\documentclass[prc,twocolumn,floatfix,groupedaddress,nofootinbib,preprintnumbers,amsmath,amssymb,amsfonts,superscriptaddress,widetable,href]{revtex4}
\usepackage{amsmath,amssymb,amsfonts}
\usepackage{hyperref}
\usepackage{ulem}
\usepackage{graphicx}
\usepackage{color}
\usepackage{longtable}

\allowdisplaybreaks

\begin{document}

\title{Structure and composition of inner crust of neutron stars from Gogny interactions}
\author{C. Mondal}
\email{chiranjib.mondal@icc.ub.edu}
\address{Departament de F{\'i}sica Qu{\'a}ntica i Astrof{\'i}sica and Institut de Ci{\`e}ncies del Cosmos (ICCUB),
Facultat de F{\'i}sica, Universitat de Barcelona, Mart{\'i} i Franqu{\`e}s 1, E-08028 Barcelona, Spain}
\author{X. Vi{\~n}as}
\email{xavier@fqa.ub.edu}
\address{Departament de F{\'i}sica Qu{\'a}ntica i Astrof{\'i}sica and Institut de Ci{\`e}ncies del Cosmos (ICCUB),
Facultat de F{\'i}sica, Universitat de Barcelona, Mart{\'i} i Franqu{\`e}s 1, E-08028 Barcelona, Spain}
\author{M. Centelles}
\email{mariocentelles@ub.edu}
\address{Departament de F{\'i}sica Qu{\'a}ntica i Astrof{\'i}sica and Institut de Ci{\`e}ncies del Cosmos (ICCUB),
Facultat de F{\'i}sica, Universitat de Barcelona, Mart{\'i} i Franqu{\`e}s 1, E-08028 Barcelona, Spain}
\author{J.N. De}
\email{jn.de@saha.ac.in}
\address{Saha Institute of Nuclear Physics, 1/AF Bidhannagar, Kolkata 700064}

\begin{abstract}

The detailed knowledge of the inner crust properties of neutron stars might be important to explain 
different phenomena such as pulsar glitches or the possibility of an {\it r-process} site in 
neutron star mergers. It has been shown in the literature that quantal 
effects like shell correction or pairing may play a relevant role to determine 
the composition of the inner crust of the neutron star. In this paper we construct 
the equation of state of the inner crust 
using the finite-range Gogny interactions, where the mean field and 
the pairing field are calculated with same interaction. We have used the semiclassical Variational 
Wigner-Kirkwood method along with shell and pairing corrections calculated with the Strutinsky 
integral method and the BCS approximation, respectively. Our results are 
compared with those of some popular models from the literature. 
We report a unified equation of state of the inner crust and core computed with 
the D1M* Gogny force, which was specifically fabricated for astrophysical calculations.

\end{abstract}
\maketitle
\section{Introduction}
The structure of a neutron star (NS) can be described mainly by a homogeneous core 
encompassed by two inhomogeneous concentric shells \cite{Haensel07}. The outermost region, 
called the ``outer-crust", is formed by a lattice of neutron-rich nuclei 
permeated by an electron gas. 
The Equation of State (EoS) in this region is mainly determined by nuclear masses, 
which can be taken from the experiment or, when unknown, 
from successful mass models. When the density reaches a value $\sim$ 
$0.0003$ fm$^{-3}$ the neutrons start to drip from the nuclei \cite{Baym71a,Haensel07,Chamel15}, 
the lattice structure still remains but now permeated by neutron and electron gases. 
This region is called 
the ``inner-crust", which further transforms into a homogeneous core around 
a density $\sim$ $0.08$ fm$^{-3}$. It has been suggested 
that near the transition to the core in the bottom layers of the inner crust 
matter may arrange in geometrical structures different from the spherical configuration in order 
to reduce the Coulomb energy. These structures may exist in various shapes 
embedded in a neutron and electron fluid giving rise to the so-called ``pasta-phase"
(see Chapter 3 of Ref. \cite{Haensel07} and references therein for comprehensive details). 
The presence of the free neutron gas lying in the continuum makes a Hartree-Fock 
(HF) calculation in this region of the NS very complicated. 
Since the seminal paper of Negele and Vautherin \cite{Negele73}, there are 
several HF calculations performed in the inner crust of NSs within the 
framework of the spherical Wigner-Seitz (WS) approximation \cite{Pizzochero02,Sandulescu04,
Baldo07,Grill11,Pastore11,Than11,Pastore15,Carreau19}, which are mainly devoted to study 
the superfluid properties of the star. 
More sophisticated three-dimensional HF calculations, often at finite temperature 
and fixed proton fraction, have recently been carried out \cite{Gogelein07,Newton09,
Pais12,Schuetrumpf14,Fattoyev17,Schuetrumpf19}. Also 
calculations based on Monte-Carlo techniques and molecular 
dynamics calculations, which do not impose periodicity or symmetry 
of the system unlike the WS approximation, have been reported in the literature 
\cite{Horowitz04,Watanabe09,Schneider13,Horowitz15}. However, these calculations usually 
do not provide the complete EoS in the inner crust. Thus, simplified methods of semiclassical 
type, based on the Thomas-Fermi approximation with non-relativistic 
\cite{Oyamatsu93,Gogelein07,Onsi08,Pearson12,Sharma15,Martin15,Lim17,Pearson18,Pearson19} 
or relativistic \cite{Cheng97,Shen98a,Shen98b,Grill12,Grill14,Bao15} interactions as well as the Compressible Liquid Drop 
Model (CLDM) calculations \cite{Lattimer91,Douchin01}, are often used for systematic studies 
of the inner crust.

Although the mass and the thickness of the NS crust is a small fraction of their 
total values, which are basically determined by the core, the crust has a 
relevant role in some observed astrophysical phenomena such as pulsar 
glitches, quasi-periodic oscillations in soft gamma ray repeaters, thermal 
relaxation in soft X-ray transients, etc \cite{Haensel07,Chamel08,Piro05,Strohmayer06,
Steiner09,Sotani12,Newton12,Piekarewicz14}. The crust might also be an {\it r-process} 
site in NS mergers \cite{Lattimer77,Meyer89,Freiburghaus99,Goriely05a}.
Therefore, it is very 
important to draw a clear picture of the structure and composition of the 
crust. A large number of theoretical studies of the inner crust of NSs have been 
carried out with Skyrme interactions or Relativistic Mean Field parametrizations. 
In this work we intend to do this analysis using the finite-range Gogny 
interactions \cite{Decharge80,Berger91}. The reason behind this is two-fold. In one hand, we wanted to 
extend the advantage of using newly proposed Gogny forces \cite{Gonzalez17,
Gonzalez18} which are able to predict maximum masses of NSs about $2M_{\odot}$, 
in agreement with the observational data \cite{Demorest10,Antoniadis13,Cromartie19}. 
On the other hand, unlike Skyrme and RMF interactions, Gogny forces describe simultaneously 
the mean-field and the pairing field, which may have some impact on the 
study of the crust.

The D1 family of Gogny 
interactions consists of two finite-range terms plus a density-dependent zero-range 
contribution. In particular, the D1S interaction \cite{Berger91} has been 
used to perform large-scale Hartree-Fock-Bogoliubov (HFB) calculations \cite{Cea,Delaroche10} of 
ground-state properties of finite nuclei along the whole periodic table. The 
D1S interaction has also been widely used for dealing with fission properties 
\cite{Schunck16} and has become a benchmark for the study of deformation and pairing properties 
of finite nuclei. 
More recent Gogny parametrizations, namely D1N \cite{Chappert08} and D1M \cite{Goriely09}, 
have been proposed. They take into account the microscopic neutron matter EoS of Friedman and 
Pandharipande in the fit of their parameters, which improves the description of the 
isovector properties. In particular, the D1M interaction is able to reproduce 
more than 2000 experimental masses with a {\it rms} deviation of only 798 keV \cite{Goriely09}. 
However, in spite the success of the Gogny forces in 
describing finite nuclei, their extrapolation to the NS domain 
has been less successful. 
It has been shown earlier \cite{Sellahewa14,Gonzalez17} that  
the Tolman-Oppenheimer-Volkov (TOV) equations, which give the NS mass-radius 
relationship, have no solution using the EoS built up 
with the D1S force. If one uses the D1N Gogny force \cite{Chappert08} the maximum 
mass is around 1.2$M_{\odot}$ and with the D1M Gogny interaction \cite{Goriely09} 
it is about 1.7$M_{\odot}$. For both these cases the 
predicted maximum mass is much lower than the observed values of about 2$M_{\odot}$ \cite{Demorest10, 
Antoniadis13,Cromartie19}. In order to overcome this limitation, 
two new Gogny interactions aimed to describe NS properties have been built up 
recently. These new parametrizations, called D1M$^*$ \cite{Gonzalez18} and D1M$^{**}$ \cite{Vinas18a}, 
are obtained by modifying the D1M force in such a way that they 
reproduce finite-nuclei data with quality similar to that found using 
the D1M force and, at the same time, predict a maximum mass of NS 
of about 2$M_{\odot}$ \cite{Gonzalez18,Vinas18a}.

In the astrophysical context, our previous studies using the modified D1M forces have been mainly applied 
for describing the core of the star. In addition to the NS mass-radius 
relationship, we have also studied the moment of inertia \cite{Gonzalez17,Gonzalez18} as well as 
the tidal deformability \cite{Vinas18b,Lourenco20}. We have also performed a detailed analysis 
of the crust-core transition using both the thermodynamical and dynamical 
methods \cite{Gonzalez19}. Some of these calculations need the knowledge of the EoS 
in the crust region. When necessary, and due to the lack of the crustal EoS 
computed with the Gogny forces, we have used a polytropic form of 
the type $P=a+b\epsilon^{4/3}$, where $P$ is the pressure and $\epsilon$ is 
the mass-energy density \cite{Gonzalez18,Vinas18b}. The parameters $a$ and $b$ are determined
in  such a way that the pressure be continuous at the inner crust-core and inner-outer crust transition
points. 
Presently, our aim is to establish the EoS in the inner crust with the 
same Gogny interaction used to describe the core.

In our calculations of the inner crust we use the 
WS approximation, which assumes that the space can be described by non-interacting 
electrically neutral cells containing each one a single nuclear cluster 
surrounded by electron  and neutron gases. We restrict ourselves to spherical  
nuclear clusters, which smoothly transform to homogeneous core at the 
transition density without the pasta phases. There are two reasons for 
that. Firstly, considering only spherical nuclei, quantal calculations 
including pairing correlations can be performed in a rather simple way. Secondly, 
as discussed e.g. in Ref. \cite{Sharma15,Chamel15}, having spherical 
nuclei or pasta structures in the region of the inner crust close to the 
core-crust transition density has very little impact on the EoS of the NS. 
Another simplification used in many  
calculations of the inner crust of NSs is to perform semiclassical calculations 
of Thomas-Fermi type in the representative WS cell assuming a mixture of 
neutrons, protons and electrons in charge and beta equilibrium (see Ref. 
\cite{Sharma15} for a detailed discussion). Quantal effects, mainly proton shell corrections 
and proton pairing correlations, can be added in a perturbative way in a 
microscopic-macroscopic (Mic-Mac) description of the inner crust of the NS. In particular, large-scale 
calculations of this type have been carried out by the Brussels-Montreal 
group \cite{Pearson12,Onsi08,Fantina13,Potekhin13,Pearson15,Pearson18} using the BSk family of Skyrme forces 
\cite{Goriely10}. 
The WS approximation induces, however, spurious shell effects in the spectrum of 
unbound neutrons \cite{Chamel07}. To avoid 
this difficulty, as was also done in the pioneering paper of Negele and 
Vautherin \cite{Negele73}, one can neglect the neutron shell effects, which 
are much smaller than the proton shell correction \cite{Chamel07}. 
Regarding pairing correlations,
we must emphasize the advantage of using Gogny forces, in which pairing is treated with the 
same interaction employed to describe the mean field.

Because of the reasons above we have chosen a Mic-Mac 
approach to establish the EoS in the inner crust using 
Gogny forces. We first compute the semiclassical EoS within the so-called 
Variational Wigner-Kirkwood approach \cite{Schuck93,Centelles98b,Centelles07} using trial neutron and proton 
densities of the type proposed in Ref. \cite{Onsi08}. This calculation includes 
the $\hbar^2$-contributions to the kinetic, exchange and spin-orbit energies perturbatively
\cite{Soubbotin00,Soubbotin03}. In a second step, the proton 
shell corrections are incorporated through the so called Strutinsky Integral Method 
\cite{Onsi08}. Finally, the proton pairing correlations are calculated in the BCS 
approximation with the corresponding Gogny interaction. 

The paper is organized as follows. In the second section the basic theory 
concerning the Variational Wigner-Kirkwood approach in the case of finite-range 
interactions is discussed. In the same section, the shell correction and the 
treatment of the pairing correlations are briefly summarized. The third section 
is devoted to the discussion of the results of the inner crust obtained in this work. 
Finally, our conclusions are given in the last section. Some of the details of our 
calculations are given in the Appendices.

\section{EoS of inner crust of neutron star}\label{formalism}
The mass formula has been a very useful tool for dealing with the average 
behavior of nuclear masses from the early days of Nuclear Physics. The success 
of the mass formula lies in the fact that the quantal effects, i.e., the shell 
correction and pairing correlations, can be treated perturbatively because 
they are small as compared with the part that 
varies smoothly with mass number $A$ and atomic number $Z$. The perturbative 
treatment of the shell correction in finite nuclei was established by 
Strutinsky \cite{Strutinsky67}. 
The average part of the energy can be extracted from a quantal mean-field HFB 
calculation using the so-called Strutinsky smoothing method 
\cite{Strutinsky67}, which is a well-defined mathematical procedure that 
removes the quantal effects. However, this method, in general, is difficult to 
handle in the case of realistic mean-field potentials
that may not vanish at the edge of the WS cells.
The reason behind this difficulty is that the Strutinsky smoothing 
requires the knowledge of the single-particle spectrum at least in three 
major shells above the Fermi level. For realistic potentials this situation 
requires taking account of states in the continuum, which is difficult to 
handle in practice.

To avoid the problem related to continuum, an alternative technique consists of computing  
the average energy of a nucleus using the Wigner-Kirkwood (WK) 
$\hbar$-expansion of the one-body partition function \cite{Wigner32, 
Kirkwood33, Uhlenbeck36, Ring80, Brack97,Schuck93}, from where the smooth part of the energy 
of a system, i.e., the energy without the quantal effects, can easily be derived
avoiding the problems related to the continuum. An important property of the WK expansion 
of the energy in powers of $\hbar$ concerns its variational content. For a set of 
non-interacting fermions in an external potential, the variational solution 
that minimizes the WK energy at each order of $\hbar$, is just the WK expansion 
of the particle density at the same $\hbar$-order. 
This method of solving the variational equation by sorting out properly the 
different powers of $\hbar$-expansion is called the Variational Wigner Kirkwood 
(VWK) theory \cite{Schuck93, Centelles07}. It is important to point out that in this variational 
method, the VWK energy at a given order of $\hbar$ is just the sum of the 
energies up to that order calculated with the densities computed up to the previous 
order. For example, calculation of the VWK energy up to $\hbar^2$-order i.e. 
sum of the variational Thomas-Fermi (TF) energy ($\hbar^0$ order) and the 
perturbative $\hbar^2$ contribution, only requires the explicit knowledge of 
the TF densities. 

\subsection{Variational Wigner-Kirkwood method for finite nuclei with the Gogny 
force}\label{vwk}
The Gogny force of the D1 family including the spin-orbit interaction 
can be written as,
\begin{eqnarray}\label{VGogny}
  V (\vec{r}_1 , \vec{r}_2) &=& \sum_{i=1,2}
  \big( W_i + B_i P_{\sigma} - H_i P_{\tau} - M_i P_{\sigma}P_{\tau}\big)
   e^{-\frac{r^2}{\mu_i^{2}}} \nonumber\\
  && + t_3 (1+ x_3 P^\sigma) \rho^\alpha (\vec{R}) \delta (\vec{r}) 
	\nonumber \\
&& + i W_{LS} (\vec{\sigma}_1 + \vec{\sigma}_2) (\vec{k'} \times \delta (\vec{r}) 
	\vec{k}),
\end{eqnarray}
where ${\vec r}={\vec r_1}-{\vec r_2}$ and ${\vec R}=({\vec r_1}+{\vec r_2})/2$ are 
the relative and the center-of-mass coordinates, and $\mu_1 \simeq$0.5-0.7 fm 
and $\mu_2 \simeq$1.2 fm are the ranges of the two Gaussian form factors, which 
simulate the short- and long-range components of the force, respectively. The 
last term in Eq. (\ref{VGogny}) is the spin-orbit interaction, which is 
of zero range with strength $W_{LS}$ as in the case of Skyrme forces. The 
quantity $\vec{k}=(\vec{\nabla_1}-\vec{\nabla_2})/2i$ is the relative 
momentum and $\vec{k'}$ its complex conjugate.

The total energy of a nucleus in the VWK approximation can be obtained starting 
from the interaction given in Eq. (\ref{VGogny}) and using the corresponding Extended 
Thomas-Fermi density matrix \cite{Soubbotin00} (see also Appendix \ref{appendix1} for more details), 
which allows to write the VWK energy upto $\hbar^2$ order as \cite{Soubbotin00,Gridnev17}: 
\begin{eqnarray}\label{EGogny}
	E_{\text{VWK}}&=&E_{\text{VWK},0}+E_{\text{VWK},2}=\int d\vec{R}\mathcal{H}\nonumber\\
&=&\int d\vec{R}\left[\mathcal{H}_{kin}+\mathcal{H}_{dir}^{nucl}+
\mathcal{H}_{exch}^{nucl}+\mathcal{H}_{coul}+\mathcal{H}_{SO}\right]\nonumber\\
&=&\int d\vec{R}\left[\mathcal{H}_{kin,0}+\mathcal{H}_{dir}^{nucl}+
\mathcal{H}_{exch,0}^{nucl}+\mathcal{H}_{coul}\right]\nonumber\\
&+&\int d\vec{R}\left[\mathcal{H}_{kin,2}+\mathcal{H}_{exch,2}^{nucl}+
\mathcal{H}_{SO}\right],
\end{eqnarray}
where we have split the energy into a TF ($\hbar^0$) part and a $\hbar^2$ correction. 
One can see that the $\hbar^2$-corrections enter into the kinetic energy 
and spin-orbit energy densities $\mathcal{H}_{kin}$ and $\mathcal{H}_{SO}$, respectively,
as it happens for zero-range forces \cite{Brack97}, and also in the exchange 
energy due to the finite range of the force. 

In order to find the semiclassical energy and the density profiles in the VWK 
approach up to $\hbar^2$ order, one needs to solve first the equation of motion for the TF problem,
\begin{eqnarray}\label{eom}
	{\delta\over{\delta\rho_q}}\left[E_{\text{VWK},0}-\mu_q\int\rho_q(\vec{R})d
\vec{R}\right]=0,
\end{eqnarray}
where $\mu_q$ is the chemical potential with $q=n$ for neutrons and $p$ for 
protons that ensures the right number of particles. Using the solutions of 
Eq. (\ref{eom}) one can calculate both the 
TF ($E_{\text{VWK},0}$) as well as the $\hbar^2$ correction ($E_{\text{VWK},2}$) of the VWK 
energy. In the present work, however, instead of solving the full variational  
equation (Eq.(\ref{eom})), we perform a restricted variational calculation minimizing 
the TF energy using trial densities of Fermi type. This technique has been 
successfully applied in many finite nuclei calculations with Skyrme forces 
\cite{Brack97}. It is shown in Ref. \cite{Centelles90} that the differences 
between the semiclassical energies obtained either by solving self-consistently 
the equations of motion or by means of a restricted variational approach are 
very small. This fact justifies the use of this simpler technique, which, 
in addition, is still more stable numerically. In the case of finite nuclei 
the profile of the trial neutron and proton density functions used to minimize 
the energy (Eq. (\ref{EGogny})) is chosen as a Fermi distribution for each 
kind of particles, 
\begin{eqnarray}\label{fermi}
\rho_q(r)=\frac{\rho_{0,q}}{1+{\rm exp}\left({{r-C_q}\over{a_q}}\right)},
\end{eqnarray}
where the radii $C_q$ and diffuseness $a_q$ are the variational parameters
and the strengths $\rho_{0,q}$ are determined by normalization to the number 
of particles of each type.
\subsection{Restoring quantal effects: Shell and Pairing corrections}
\label{shellpair}
It has been mentioned in the previous subsection that the VWK method provides 
the average energy of the nucleus. To obtain the quantal energy in a 
Mic-Mac  approach one needs to add perturbatively the shell 
effects and, in the case of open shell nuclei, also incorporate the energy due to the 
pairing correlations. To compute the shell correction we use the 
so-called Strutinsky integral method \cite{Pearson12}, which states that 
for each type of particles the shell correction can be estimated as the 
difference between the quantal and semiclassical energies in the corresponding 
semiclassical single-particle potential treated as an external one:
\begin{eqnarray}\label{Eshell}
E_q^{shell}=\sum_i \widetilde\epsilon_{i,q}-\int d\vec{R}\left[{{\hbar^2}\over
{2\widetilde m_q^*}}\widetilde\tau_q+\widetilde\rho_q\widetilde U_q+\widetilde{\vec{J}}_q\cdot
\widetilde{\vec{W}}_q\right],
\end{eqnarray}
where, the single particle energies $\widetilde\epsilon_{i,q}$ are the eigenvalues of 
the Schr{\"o}dinger equation as follows, 
\begin{eqnarray}\label{Schrodinger}
\left[-{\vec{\nabla}}{\hbar^2\over{2\widetilde m_q^*}}{\vec{\nabla}}
+\widetilde U_q-i\widetilde{\vec{W}}_q\cdot({\vec{\nabla}}\times\vec\sigma)
\right]\phi_{i,q}=\widetilde\epsilon_{i,q}\phi_{i,q},
\end{eqnarray}
for each type of particles ($q=n,\ p$).

In Eqs. (\ref{Eshell}) and (\ref{Schrodinger}) the quantities $\widetilde\rho_q$, 
$\widetilde\tau_q$ and $\widetilde{\vec{J}}_q$ are the particle, kinetic energy
and spin densities, respectively. The effective mass, central and
spin-orbit potentials are denoted by $\widetilde m_q^*$, $\widetilde U_q$ and
$\widetilde{\vec{W}}_q$, respectively. All these smooth densities and fields
mentioned above are evaluated with the semiclassical solutions of the restricted
variational approach applied to the TF energy. It is also worthwhile to note
that the Schr{\"o}dinger equation (\ref{Schrodinger}) is obtained from the
quasilocal reduction of the energy density associated to a finite range 
forces in the framework of the non-local Density Functional Theory (DFT)
\cite{Soubbotin03} as explained in Appendix A.

Pairing correlations are taken into account at the BCS level in a
perturbative way using the single particle levels obtained from Eq.
(\ref{Schrodinger}). This implies that the single particle gaps obey, for
each type of particles, the following set of coupled gap equations 
\begin{eqnarray}\label{Gap}
\Delta_i=-\sum_k v_{i\bar i, k\bar k}^{pair}\ \ {\Delta_k\over{2E_k}},
\end{eqnarray}
where, the indices $i\equiv nlj$ and $k\equiv n^\prime l^\prime j^\prime$
denote the corresponding single particle levels, $v_{i\bar i, k\bar
k}^{pair}$ are the pairing matrix elements calculated with the Gogny force (for
more details see Appendix B of \cite{Than11}) and 
$E_k=\sqrt{(\widetilde\epsilon_{k}-\mu)^2+\Delta_k^2}$ are the quasiparticle 
energies. Once the gap equations are solved, the pairing energy and the 
occupation number of each level can be computed as,
\begin{eqnarray}\label{Epair}
E^{pair}=-{1\over 4}\sum_k {\Delta_k^2\over{E_k}}
\end{eqnarray}
and 
\begin{eqnarray}\label{Occupation}
\widetilde n_k^2={1\over 2}\left[1-{{\widetilde\epsilon_k-\mu}\over E_k}\right],
\end{eqnarray}
respectively.

Notice that the shell and pairing corrections are calculated simultaneously, 
implying that the occupation probabilities (Eq. (\ref{Occupation})) need to
be included in the calculation of the shell correction and the total energy
of a nucleus in this case is given by 
\begin{eqnarray}\label{Etot}
	E=E_{\text{VWK}}+\sum_q\left[E_q^{shell}+E_q^{pair}\right].
\end{eqnarray}

\subsection{VWK method with shell and pairing corrections in 
Wigner-Seitz cells}\label{vwkws}
To deal with the inner crust of neutron star we have adopted the WS 
approximation. As mentioned before, we do not delve into the possibilities of 
pasta phase in the inner crust matter in the present calculation. We restrict ourselves to a spherical 
WS approximation to describe the building blocks of the inner crust. For a 
given density in the WS cell, we look for the configuration ($N,Z$) of the WS cell with a given number 
of protons  and electrons that fulfills the $\beta$-equilibrium condition given by, 
\begin{eqnarray}\label{betaequilibrium}
\mu_n=\mu_p+\mu_e,
\end{eqnarray}
where, $\mu_n$, $\mu_p$ and $\mu_e$ are the neutron, proton and electron 
chemical potentials, respectively. For a given average density and proton number 
we start with a trial number of neutrons and compute the difference 
$d\beta=\mu_n-\mu_p-\mu_e$. If this quantity does not vanish, we change 
the number of neutrons governed by the deviation of 
$d\beta$ from zero using Newton-Raphson method. This eventually changes the
size of the cell because the total number of nucleons determines 
the radius of the WS cell. This procedure is iterated till $d\beta$ approaches zero 
with some chosen accuracy. Once this configuration (neutron, proton and 
electron content) of the the WS cell is determined, the radius of the cell is 
also decided. To obtain the energy we apply the techniques similar to the ones 
used to compute the energy of finite nuclei as described in the subsection 
\ref{vwk} and Appendix \ref{appendixb}. The pressure associated to the WS cell 
is computed as explained in Ref. \cite{Pearson12}. There is, however, a small change 
in the form of the trial density for baryons used to solve the equations of 
motion (Eq. (\ref{eom})), which is taken from Ref. \cite{Pearson12}, 
\begin{eqnarray}\label{fermiws}
\rho_q(r)=\rho_{B,q}+\frac{\rho_{0,q}}{1+{\rm exp}\left\{
\left({{C_q-R_{WS}}\over{r-R_{WS}}}\right)^2-1\right\}{\rm exp}
\left({{r-C_q}\over{a_q}}\right)},\ \ \ 
\end{eqnarray}
where the first term represents a background density throughout the WS cell and 
the usual Fermi distribution has an extra damping factor modulated by the 
radius of the WS cell $R_{WS}$. This damping ensures that all the density 
derivatives vanish at the edge of the WS cell, which implies a smooth matching of 
the adjacent cells justifying the WS approximation of the inner crust. 
We recover the quantal effects of protons in terms of shell and 
pairing corrections in the same fashion 
as described in the section \ref{shellpair}. 

For a set density we repeat the procedure described above for even numbers 
of protons varying from $Z=14$ to $Z=100$. The optimal configuration (neutron, proton and electron 
content) giving the minimum energy is designated to give the 
description of the inner crust at a particular density. It is worthwhile to 
mention here that in our calculation the number of neutrons is not integer to 
achieve $\beta$-equilibrium inside the WS cell. The charge neutrality of 
a WS cell is maintained by considering an electron gas distributed uniformly in the 
whole cell in such a way that the number of electrons equal to that of protons in the cell. 
The contribution of the electrons to the energy of the cell is taken into account 
considering them to be ultra-relativistic and including the direct and 
exchange Coulomb energy at the Slater level.

\begin{figure}[]{}
\includegraphics[height=3.5in,width=3.2in,angle=-90]{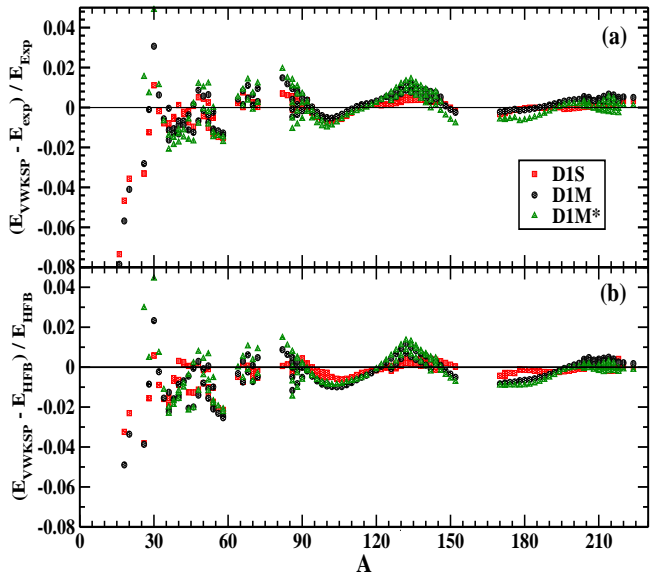}
\caption{\label{fig1}Relative difference between binding energies  calculated 
	from VWKSP approach with HFB method in harmonic oscillator basis and their experimental values 
	for $\sim 160$ even-even nuclei spread across the whole nuclear chart 
	using D1M, D1S and D1M* Gogny forces.}
\end{figure}
\section{Results and Discussions}\label{results}
\begin{figure}[]{}
\includegraphics[height=3.5in,width=3.2in,angle=-90]{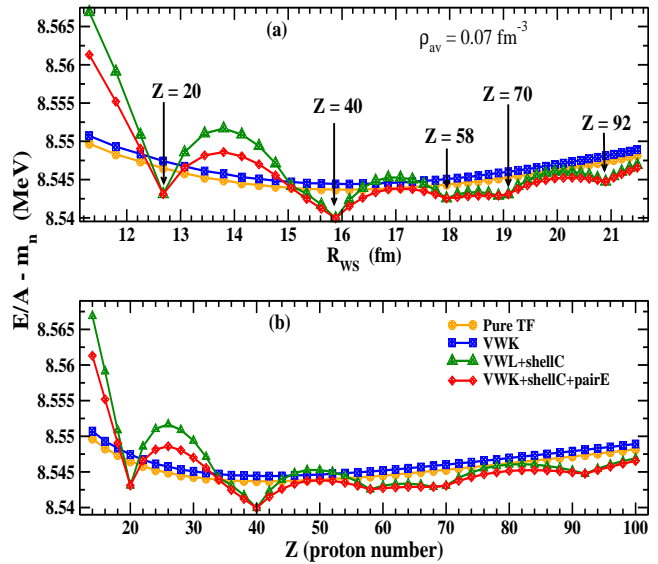}
\caption{\label{fig2}Binding energy per particle subtracted by the free nucleon mass 
	for inner crust of neutron star calculated with D1M* force as a function of 
	radius of the WS cell in panel (a) and as a function of proton number $Z$ in 
	panel (b) at average density $\rho_{\rm av}=0.07$ fm$^{-3}$. }
\end{figure}
\subsection{Finite Nuclei}
\begin{table*}[]
        \caption{Binding energies of 10 different nuclei across the nuclear chart calculated with VWKSP 
	method and compared with their experimental values along with the ones calculated with  
	Hartree-Fock-Bogoliubov (HFB) method in harmonic oscillator basis using D1M* Gogny force.}
  \label{tab1}
\begin{tabular}{ccccccccccc}
\toprule
	Nucl.($\rm _{Z}^AX_N$)& $\rm _{12}^{32}Mg_{20}$ & $\rm _{20}^{40}Ca_{20}$ & $\rm _{20}^{50}Ca_{30}$ & $\rm _{40}^{90}Zr_{50}$ & $\rm _{50}^{100}Sn_{50}$ &
	$\rm _{62}^{142}Sm_{80}$ & $\rm _{80}^{176}Hg_{96}$ & $\rm _{82}^{208}Pb_{126}$ & $\rm _{84}^{216}Po_{132}$ & $\rm _{92}^{224}U_{132}$ \\ 
Expt. & \small -249.849  & \small -342.052 & \small -427.490 & \small -783.892 & \small -824.794 &
        \small -1176.614 & \small -1369.743 & \small -1636.430 & \small -1675.904 & \small -1710.285 \\
VWKSP & \small -251.665  & \small -338.010 & \small -429.318 & \small -784.834 & \small -818.355 &
        \small -1174.188 & \small -1357.301 & \small -1636.241 & \small -1679.186 & \small -1708.199 \\
HFB & \small -248.848  & \small -342.546 & \small -424.996 & \small -783.138 & \small -826.542 &
        \small -1174.549 & \small -1364.428 & \small -1636.729 & \small -1671.738 & \small -1706.392 \\
\toprule
 \end{tabular}
\end{table*}
\begin{figure*}[]{}
\includegraphics[height=6.0in,width=4.0in,angle=-90]{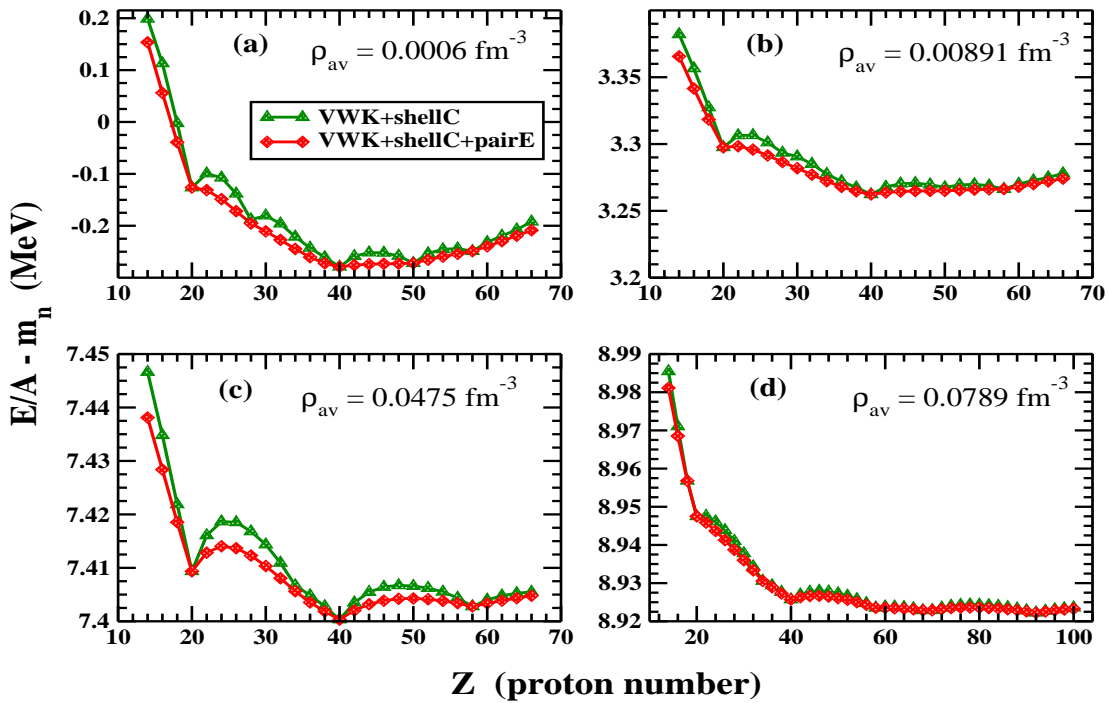}
\caption{\label{fig3}Binding energy per particle for inner crust of neutron 
	star subtracted by free nucleon mass calculated at different average densities mentioned in different 
	panels using VWK approach with shell correction and pairing are plotted as a function of proton number 
	$Z$ for D1M* Gogny force in the spherical WS approximation.}
\end{figure*}
Before applying the VWK method with shell and pairing correction (VWKSP) to 
the inner crust, we take recourse to testing it for finite nuclei. For this 
purpose we have used three different Gogny forces, namely, D1S, D1M and D1M$^*$. 
We want to recall here that all three forces mentioned previously 
belong to D1 family of Gogny forces with the interaction given by Eq. (\ref{VGogny}). 
As it was pointed out before, the new D1M$^*$ force was built with the aim to
be able to predict a maximum NS mass of about 2$M_{\odot}$ and, at the same time, describe
finite nuclei with a quality similar to that found using the D1M interaction.
To this end, starting with the D1M force, the eight parameters that determine the
finite-range part of the force were modified as follows. The saturation density, binding energy per
nucleon, incompressibility coefficient and effective mass in symmetric nuclear matter, as well
as the symmetry energy at a subsaturation density ($\rho=0.1$ fm$^{-3}$) in the isovector
sector, were kept fixed to the values predicted by the D1M force. The two combinations
of $W_i,B_i,H_i$ and $M_i\ (i=1,2)$, which determine the strength of the pairing force in
the $S=0,T=1$ channel, were also kept at the same value as in the original D1M force. In this way seven
out of the eight parameters were determined. The last parameter, chosen to be $B_1$,
was used to modify the slope
of the symmetry energy at saturation, which in turn determines the maximum mass of the
neutron star. Finally, the strength of the zero-range part of the force and the
spin-orbit strength in Eq.(\ref{VGogny}) were fine tuned for improving the description of finite
nuclei (see Ref.\cite{Gonzalez18} for further details).

We have calculated the binding energies of a set of even-even spherical nuclei and 
compared them with their experimental values in Fig \ref{fig1}(a). In this figure 
we plot the relative difference in the binding energy for $\sim 160$ even-even nuclei 
spread across the entire nuclear chart, calculated with the VWKSP method using the 
Gogny forces mentioned before from their experimental values. One can observe that 
the differences for the lighter nuclei are on the higher side compared to those for 
heavier nuclei. This is a typical feature of the mean-field approximation, which 
works better with a relatively large number of particles in the system. From this 
panel we see that for mass numbers larger than $A\sim 30$ the VWKSP energies are scattered 
around the experimental values reproducing them within a 2$\%$ of accuracy. In order 
to validate our VWKSP method to predict binding energies, we compare our results 
with the ones obtained using the standard HFB method, 
which is the benchmark for the theoretical ground-state binding energies with effective forces. 
The relative differences between the HFB and VWKSP methods are plotted in Fig. 
\ref{fig1}(b).
They differences are quite small, less than 2$\%$ almost throughout the entire nuclear 
chart apart from few nuclei in the lighter region. In Table \ref{tab1}, a representative 
subset of these nuclei is provided with their binding energies 
computed with VWKSP method using the D1M* interaction and their comparison with the experimental values as well 
as with the corresponding HFB results. Overall the agreement is 
satisfactory. These tests in the finite nuclear sector
provide us the much needed confidence to apply VWKSP method for the case of
inner crust of NS.
\begin{figure*}[]{}
\includegraphics[height=6.0in,width=4.0in,angle=-90]{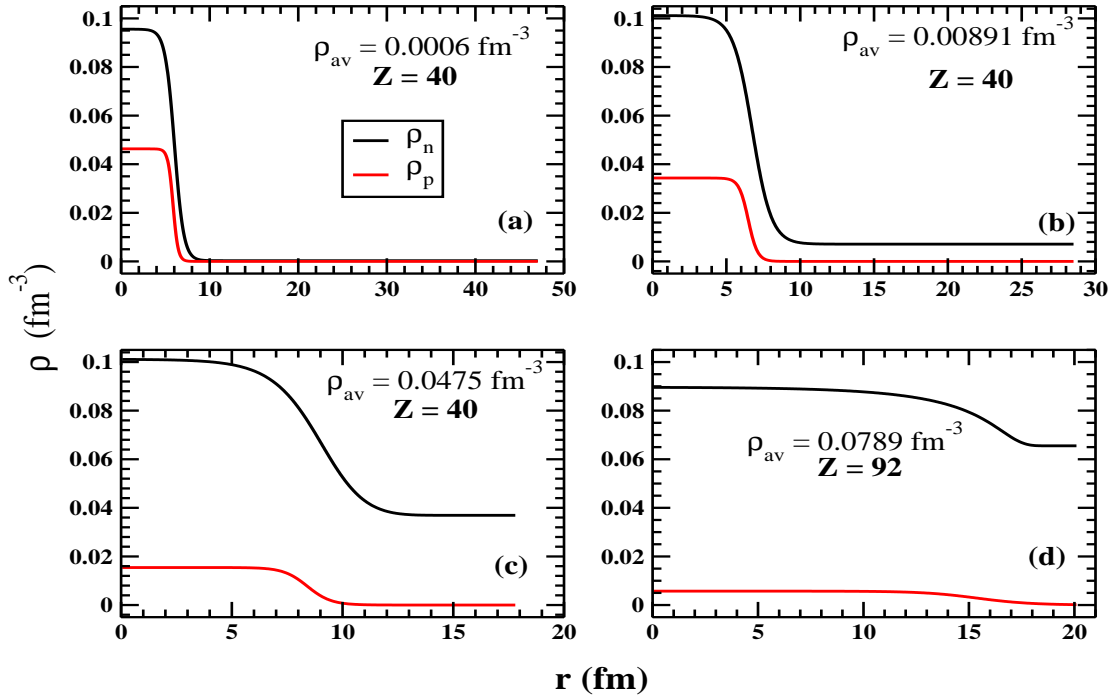}
	\caption{\label{fig4}Same as Fig. \ref{fig3} but for neutron and proton 
	density profiles inside the WS cell.}
\end{figure*}
\subsection{Inner crust of Neutron Stars}
We start the calculation of inner crust by choosing different average densities $\rho_{\rm av}$
ranging from 0.0003 fm$^{-3}$ to 0.08 fm$^{-3}$. Some of the choices are 
taken from the article by Negele and Vautherin \cite{Negele73}. 
For each average density we search the optimal configuration, i.e., neutron and 
proton numbers which minimize the energy per particle in the WS cell using 
the VWKSP method. To this end, we proceed in two steps. In the first one, for 
an even proton number $Z$ we determine the neutron number $N$ from the 
$\beta$-equilibrium condition (Eq. \ref{betaequilibrium}), which also determines the radius 
of WS cell. In the second step we scan the energy per particle as a function of 
$Z$ to obtain the global minimum. 
In panel (a) of Fig. \ref{fig2}, we plot the energy per particle (subtracted by free nucleon mass, $m_n=939$ MeV) 
as a function of radius of the WS cell, $R_{WS}$, for an average density $\rho_{\rm av}=0.07$ 
fm$^{-3}$ using the D1M* interaction. We plot the same quantity as a function of proton 
number $Z$ in Fig. \ref{fig2}(b).
The orange lines with circles represent the pure 
Thomas-Fermi (TF) calculation. Being a semiclassical calculation in 
nature, the energy per particle varies smoothly with $N$ and $Z$ in this case. 
The blue lines with the squares represent the VWK calculation, 
where the $\hbar^2$-correction is added to the TF energy perturbatively. 
The green lines with triangles correspond to the results
when the shell correction is added further, perturbatively. Clearly this 
quantal effect makes the graphs unsmooth in nature and distinct local 
minima appear at conventional magic numbers like $Z=20$ or 40, but there 
also appear few other magic numbers at $Z=58,70,92$ and 138 (the latest not shown in 
Fig. \ref{fig2}). The red lines 
with diamonds correspond to added pairing energy on top of the 
shell correction. This somewhat smoothens the lines compared to those 
only with shell correction. The global minimum appears at $Z=40$ after incorporation 
of all the corrections, which describes the optimal configuration of the crust at 
$\rho_{\rm av}=0.07$ fm$^{-3}$ calculated with the D1M* interaction. 
This systematic study demonstrates the importance of quantal 
effects to determine the configuration of the 
inner crust of neutron star. Notice that the local minima, which correspond to 
magic proton numbers, are unaffected by pairing correlations as they have 
an impact only on the open-shell nuclei lying in between the minima. 
One can notice from Fig. \ref{fig2} that with increase in proton number 
the size of the WS cell does not increase linearly. 
When the proton number increases in a WS cell of given density, the neutron 
number also increases to maintain the $\beta$-equilibrium and, therefore, 
the size of the cell, which is determined by the mass number, also grows.

\begin{figure*}[]{}
\includegraphics[height=6.0in,width=4.0in,angle=-90]{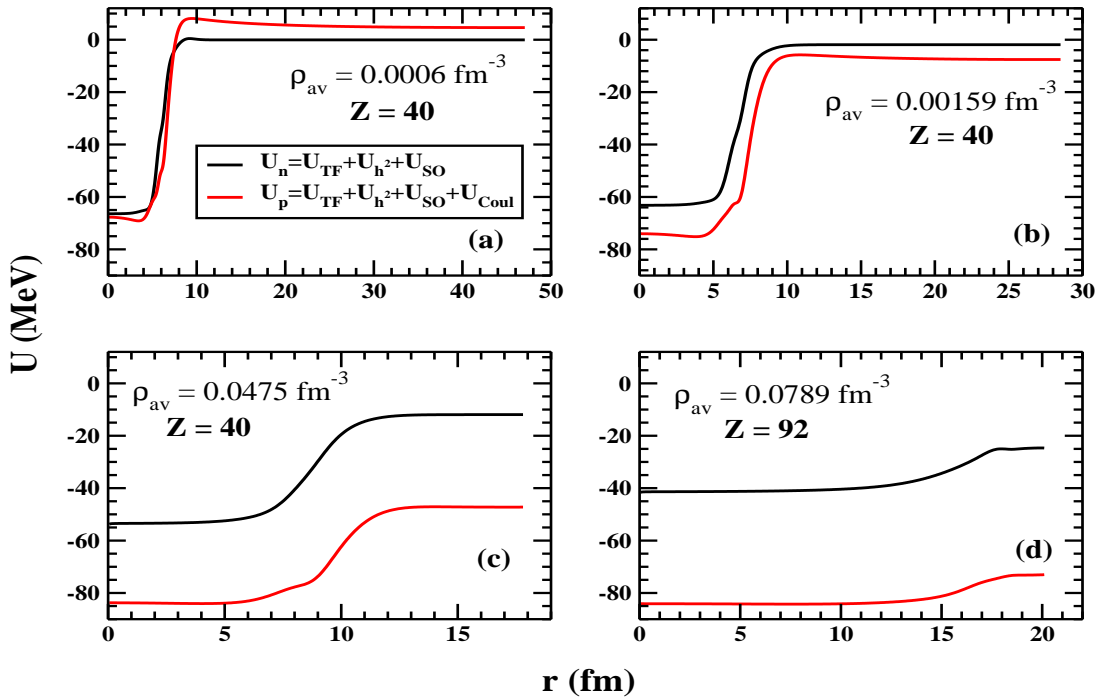}
	\caption{\label{fig5}Same as Fig. \ref{fig3} but for neutron and proton 
	potentials inside the WS cell.}
\end{figure*}
In Fig. \ref{fig3} we plot the energy per particle subtracted by the 
free nucleon mass as a function of proton number $Z$ for four different 
$\rho_{\rm av}$ values using the D1M* interaction. The green lines with the triangles 
correspond to VWK calculation along with shell correction added perturbatively and 
the red lines with diamonds correspond to VWKSP calculation, which also includes pairing 
effects. The general 
feature one can observe here is that addition of pairing somewhat smoothens 
the quantal effects throughout the average density range considered in the 
four panels. At $\rho_{\rm av}=0.0006$ fm$^{-3}$ (see Fig. \ref{fig3}(a)) only with shell correction, 
one can observe the appearance of conventional magic numbers like $Z=20,28,40,
50$. However, there is also a hint of local minimum at $Z=58$. Due to the presence 
of the neutron gas, the appearance of the magic numbers similar to those of finite nuclei gets blurred 
by the time one reaches a higher average density (for example $\rho_{\rm av}=0.00891$ fm$^{-3}$ in Fig. \ref{fig3}(b)) 
and the appearance of magic numbers like $Z=28$ or 50 gets completely washed 
away when one reaches even higher average density (say, $\rho_{\rm av}=0.0475$ fm$^{-3}$ depicted in Fig. \ref{fig3}(c)). The global minima 
appear at $Z=40$ for average densities upto $\sim 0.08$ fm$^{-3}$, where the 
minimum shifts to $Z=92$ (see Fig. \ref{fig3}(d)). One can also notice 
that the effects of shell correction get diluted when one shifts to 
higher average densities (notice that the vertical scales of the panels of 
Fig. \ref{fig3} are different for different panels).
Overall, the inclusion of quantal effects in the energy is quite crucial 
for the inner crust of neutron star. 
We must mention here is that, as a general feature, the semiclassical energy 
per particle in a WS cell varies very slowly with the proton number. In 
spite of the larger fluctuations after incorporation of shell and pairing energy 
emerge, the values of the total energy of the system are very similar for 
varying number of protons. This implies that determination of minimal 
energy configuration corresponding to the inner crust of neutron star is a 
delicate problem from the numerical point of view. 

\begin{table*}[]
        \caption{Radius of the WS cell $R_{WS}$, its corresponding neutron number $N$, proton number $Z$, energy per particle 
	subtracted by nucleon mass $E/A-m_n$, pressure $P$, chemical potentials of neutron ($\mu_n$), proton 
	($\mu_p$) and electron ($\mu_e$) at different average densities $\rho_{\rm av}$ 
	of inner crust of NS corresponding to the minimum energy configuration for D1S, D1M and 
	D1M* Gogny interactions. We have used $m_n=939$ MeV.}
  \label{tab2}
\begin{tabular}{cccccccccc}
\toprule
	 & $\rho_{\rm av}$ & $R_{WS}$ & $N$ & $Z$ & $E/A-m_n$ & $P$ & $\mu_n$ & $\mu_p$ & $\mu_e$ \\ 
	 & (fm$^{-3}$) & (fm) & & & (MeV) & (MeV fm$^{-3}$) & (MeV) & (MeV) & (MeV) \\
	\hline
D1S & 0.0004  & 50.0488 &   170.0538 &  40  &  -0.85386   &    0.00052 &     0.5105    &   -25.0660   &    25.5769 \\
	& 0.0006  & 48.1364 &   240.3249 &  40  &  -0.36003   &    0.00067 &     0.8037    &   -25.7910   &    26.5951 \\
	& 0.000879& 46.1657 &   322.2734 &  40  &   0.05280   &    0.00092 &     1.1274    &   -26.6055   &    27.7329 \\
	& 0.00159 & 42.6967 &   478.4050 &  40  &   0.66983   &    0.00170 &     1.7578    &   -28.2341   &    29.9917 \\
	& 0.00373 & 36.8371 &   741.0056 &  40  &   1.67778   &    0.00505 &     3.0441    &   -31.7328   &    34.7766 \\
	& 0.00577 & 33.5312 &   871.1931 &  40  &   2.31735   &    0.00926 &     3.9311    &   -34.2858   &    38.2171 \\
	& 0.00891 & 30.1276 &   980.6055 &  40  &   3.07788   &    0.01714 &     5.0082    &   -37.5434   &    42.5520 \\
	& 0.0204  & 23.6452 &  1089.6526 &  40  &   4.98537   &    0.05594 &     7.7295    &   -46.5564   &    54.2863 \\
	& 0.03    & 20.7605 &  1084.4102 &  40  &   6.13499   &    0.09690 &     9.3633    &   -52.5260   &    61.8894 \\
	& 0.0475  & 17.5449 &  1034.5676 &  40  &   7.77519   &    0.18563 &    11.6701    &   -61.7009   &    73.3706 \\
	& 0.06    & 12.7870 &   505.4683 &  20  &   8.73363   &    0.25872 &    12.9645    &   -67.4367   &    80.4012 \\
	\hline
D1M & 0.0004  & 48.7506 &   154.1277 &  40  &  -0.86383   &    0.00057 &     0.4688    &   -25.7906   &    26.2594 \\
	& 0.0006  & 46.7373 &   216.5842 &  40  &  -0.36546   &    0.00073 &     0.7844    &   -26.6088   &    27.3930 \\
	& 0.000879& 44.6674 &   288.1330 &  40  &   0.05627   &    0.00100 &     1.1325    &   -27.5327   &    28.6653 \\
	& 0.00159 & 41.0245 &   419.8488 &  40  &   0.69759   &    0.00185 &     1.8132    &   -29.4046   &    31.2174 \\
	& 0.00373 & 34.9033 &   624.3484 &  40  &   1.76817   &    0.00548 &     3.2011    &   -33.5087   &    36.7096 \\
	& 0.00577 & 31.5165 &   716.6228 &  40  &   2.45137   &    0.00995 &     4.1402    &   -36.5290   &    40.6694 \\
	& 0.00891 & 28.1298 &   790.7460 &  40  &   3.25485   &    0.01801 &     5.2393    &   -40.3485   &    45.5880 \\
	& 0.0204  & 25.1930 &  1308.3346 &  58  &   5.16037   &    0.05291 &     7.7310    &   -49.7482   &    57.4793 \\
	& 0.03    & 22.3713 &  1348.9628 &  58  &   6.19822   &    0.08396 &     8.9590    &   -55.8469   &    64.8064 \\
	& 0.0475  & 19.4933 &  1415.8068 &  58  &   7.52030   &    0.14037 &    10.4099    &   -64.1245   &    74.5343 \\
	& 0.06    & 18.2139 &  1460.6204 &  58  &   8.21455   &    0.18126 &    11.1399    &   -68.7630   &    79.9025 \\
	& 0.0789  & 19.5763 &  2387.4758 &  92  &   9.04198   &    0.24948 &    12.0330    &   -74.6456   &    86.6787 \\
	\hline
D1M*& 0.0004  & 49.0468 &   157.6882 &  40 &   -0.74943  &     0.00056   &   0.4855  &     -25.6148  &     26.1005 \\
	& 0.0006  & 47.0450 &   221.6858 &  40 &   -0.27892  &     0.00072   &   0.7993  &     -26.4144  &     27.2135 \\
	& 0.000879& 44.9802 &   295.0745 &  40 &    0.12384  &     0.00098   &   1.1463  &     -27.3191  &     28.4656 \\
	& 0.00159 & 41.3408 &   430.5692 &  40 &    0.74518  &     0.00183   &   1.8248  &     -29.1535  &     30.9779 \\
	& 0.00373 & 35.2332 &   643.3648 &  40 &    1.79723  &     0.00543   &   3.2033  &     -33.1619  &     36.3650 \\
	& 0.00577 & 31.8661 &   742.0830 &  40 &    2.47121  &     0.00983   &   4.1300  &     -36.0917  &     40.2217 \\
	& 0.00891 & 28.5108 &   824.9535 &  40 &    3.26236  &     0.01770   &   5.2060  &     -39.7706  &     44.9766 \\
	& 0.0204  & 22.6141 &   948.2304 &  40 &    5.12315  &     0.05114   &   7.5829  &     -49.2004  &     56.7834 \\
	& 0.03    & 20.2603 &  1005.0806 &  40 &    6.12660  &     0.08046   &   8.7544  &     -54.6898  &     63.4438 \\
	& 0.0475  & 17.8000 &  1082.1295 &  40 &    7.40033  &     0.13496   &  10.1676  &     -62.1776  &     72.3449 \\
	& 0.06    & 16.6427 &  1118.5374 &  40 &    8.07662  &     0.17845   &  10.9405  &     -66.5598  &     77.5002 \\
    & 0.07    & 15.8808 &  1134.3594 &  40 &    8.53993  &     0.22176   &  11.5152  &     -69.8479  &     81.3629 \\
    & 0.0789  & 20.0985 &  2591.2263 &  92 &    8.92225  &     0.26701   &  12.0208  &     -72.7213  &     84.7418 \\
\toprule
 \end {tabular}
\end{table*}
In Fig. \ref{fig4} we plot the density profiles for neutrons and protons 
inside the WS cell at four different $\rho_{\rm av}$ values as the ones considered in 
Fig. \ref{fig3} corresponding to their minimum energy configuration computed with the D1M* Gogny force. 
The black solid lines correspond to the neutron density profile and 
the red lines denote the ones for protons. For each average density 
we mention the number of protons corresponding to the 
minimum energy configuration described before in Fig. \ref{fig3}. 
By close inspection one can notice that at an average density as low as 
$0.0006$ fm$^{-3}$ (depicted in Fig. \ref{fig4}(a)), the density profiles for both protons and neutrons resemble 
very much to those in finite nuclei, maintaining a constant density from the center to 
certain extent and then fall down in a very short distance as compared with the size of the WS cell. 
Although not visible in Fig. \ref{fig4}(a), 
the neutron density profile maintains a feeble presence throughout the whole 
WS cell owing to the neutron gas, unlike finite nuclei where it vanishes. The proton density profile 
behaves differently and vanishes inside the WS cell, because at such low average 
densities there are no dripped protons in the inner crust.
With increasing average density of the WS 
cell, the neutron gas becomes more and more prominent. More quantitatively, if we take the case of lowest 
average density $\rho_{\rm av}=0.0006$ fm$^{-3}$ considered here, the neutron and 
proton densities remain constant, respectively, at $\sim$ 0.095 fm$^{-3}$ and $\sim$ 0.045 
fm$^{-3}$ from the center of the WS cell to about radius $r\sim 8$ fm and they fall 
almost to zero at $r\sim 10$ fm. 
For a large average density such as $0.0789$ fm$^{-3}$ (represented 
in Fig. \ref{fig4}(d)), the neutron density 
profile reaches $0.09$ fm$^{-3}$ at the center and attenuates to 0.065 fm$^{-3}$ at 
$r\sim 17$ fm. However, in this case the proton density 
remains constant at 0.01 fm$^{-3}$ starting from the center and vanishes at 
$r\sim 17$ fm. So, the WS cell is filled with dense neutron gas for this case 
and protons tend to spread in the whole WS cell. A gradual evolution can be observed through 
the intermediate average densities depicted in Fig. \ref{fig4}. 
The central value of the neutron density remains roughly constant and the 
diffuseness increases when the $\rho_{\rm av}$ in the WS cell increases. 
However, the proton densities behave  a little bit differently; the central density 
decreases and the diffuseness increases when the average density grows. 
This is because at high average densities, protons tend to maintain an almost 
uniform distribution in the whole WS cell in an attempt to reduce the Coulomb energy.
\begin{figure*}[]{}
\includegraphics[height=6.0in,width=4.0in,angle=-90]{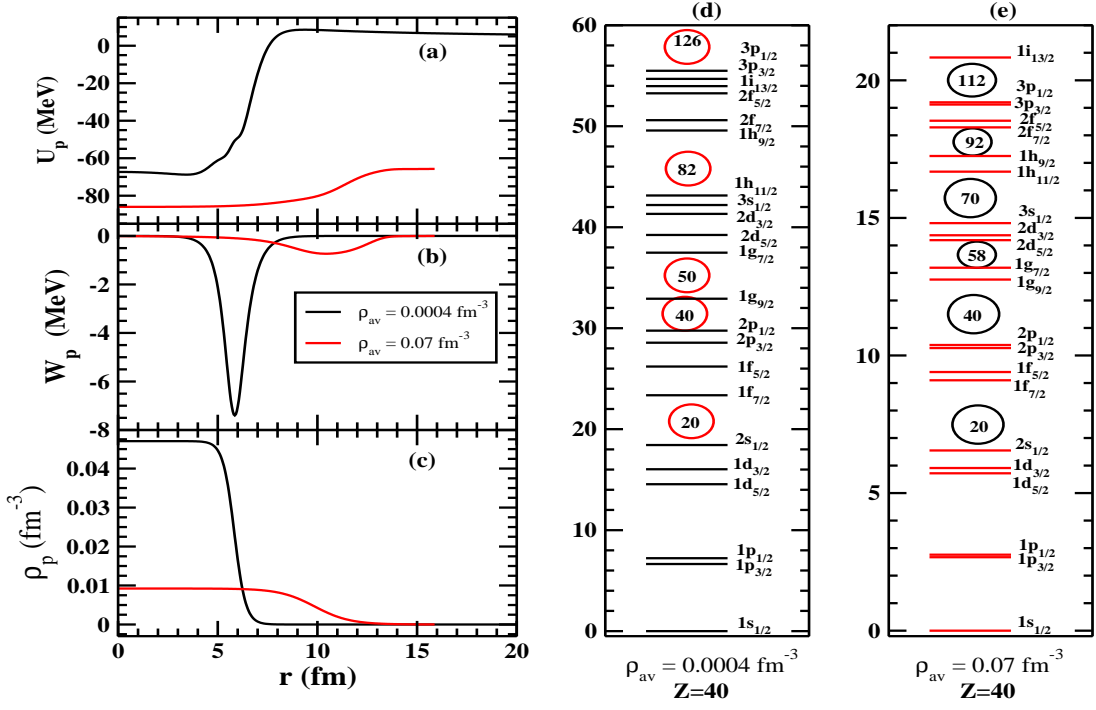}
	\caption{\label{fig6}Single-particle potential, spin-orbit potential 
	and density distributions are compared for protons in inner crust of 
	neutron star at $\rho_{\rm av}=0.0004$ fm$^{-3}$ and 0.07 fm$^{-3}$, respectively, in panels (a-c) 
	calculated with D1M* interaction. The boxes are truncated at $r=20$ fm 
	for better comparison. In panels (d-e), the single particle 
	energy levels for protons calculated with the same interaction are compared. 
	Possible appearance of magic numbers are encircled. }
\end{figure*}

In Fig. \ref{fig5} we plot the single-particle potentials for neutrons and protons inside 
the WS cell corresponding to the minimum energy configurations for the different average 
densities considered in Figs. \ref{fig3} and \ref{fig4} 
computed using the D1M* interaction. As indicated in the figure, these potentials consist 
of the self-consistent TF part plus the $\hbar^2$ and the spin-orbit contributions to the 
mean field added perturbatively (see Appendix A for further details). These proton potentials 
also include the contribution from Coulomb interaction taking into account the 
contribution of the electrons. We use this form as 
$U_q$ ($q=n$ for neutrons and $p$ for protons) of the potential in the Schr\"odinger equation
(\ref{Schrodinger}), which allows one to find the single-particle levels needed to compute the 
shell and pairing corrections to the semiclassical energy through the Strutinsky integral 
method and BCS approximation, respectively. 
These potentials, however, show trends different from the case of finite nuclei, especially in 
larger average densities. This is due to the presence of the denser neutron and electron gases, which 
deeply modify the potentials. The depths of the proton potentials decrease with 
increase in the average density of the WS cell. For example, at average 
density $\rho_{\rm av}=0.0006$ fm$^{-3}$ (in Fig. \ref{fig5}(a)), the depth of the proton potential is 
$\sim -65$ MeV and even has a Coulomb barrier of $\sim10$ MeV before attenuating 
at $r\sim 20$ fm. 
Because of the 
screening of the proton potential in the crust by electrons, the Coulomb potential 
is reduced compared to the case of terrestrial finite nuclei. 
The depth of the neutron potential is $\sim -65$ MeV 
and it goes to a constant value corresponding to the single-particle potential of
the neutron gas at $r\sim 10$ fm. On the other extreme 
of considered average density $\rho_{\rm av}=0.0789$ fm$^{-3}$ (represented in Fig. \ref{fig5}(d)), the depth of the 
proton potential is $\sim -85$ MeV at the center of the cell, which freezes 
to $\sim -70$ MeV at $r\sim 17$ fm making the effective depth of only 
about $\sim -15$ MeV. For neutrons, the potential at the center is $\sim -40$ 
MeV and attenuates to $\sim -25$ MeV at $\sim 17$ fm. Unlike finite nuclei 
this situation is very unique, which is due to the existence of the neutron 
gas that has an important impact not only on the  neutrons but also on the 
protons contained in the WS cell. On the other hand, the frozen value 
of the neutron potential is much smaller compared to those of protons. This 
indicates to the smaller chemical potential for neutrons compared to those 
for protons throughout the whole region of inner crust of neutron stars 
(see Table \ref{tab2} for further details).

In Table \ref{tab2}, we summarize the value of WS radius,
neutron number $N$ and proton number $Z$ corresponding to their 
$\beta$-equilibrium configuration, the energy per particle subtracted by the 
free nucleon mass, pressure  and the chemical potentials for neutron, proton and 
electron respectively, for all the average densities considered in the 
present work using the D1S, D1M and D1M* interactions. For the D1S interaction, apart 
from $\rho_{\rm av}=0.06$ fm$^{-3}$, for all other average densities the 
minimum energy configuration appears at $Z=40$ and it becomes $Z=20$ for 
$\rho_{\rm av}=0.06$ fm$^{-3}$. For D1M interaction over the growing 
$\rho_{\rm av}$ minimum energy configuration shifts from $Z=40$ to 
$Z=58$ at $\rho_{\rm av}=0.0204$ fm$^{-3}$ and at $\rho_{\rm av}=0.0789$ fm$^{-3}$ it appears at $Z=92$. 
For D1M* for all lower average densities the minimum energy configuration 
corresponds to $Z=40$ that shifts to $Z=92$ at $\rho_{\rm av}=0.0789$ fm$^{-3}$, 
which is almost the transition density estimated by our inner crust calculation using the 
D1M* interaction. 

\begin{figure}[]{}
\includegraphics[height=3.5in,width=3.2in,angle=-90]{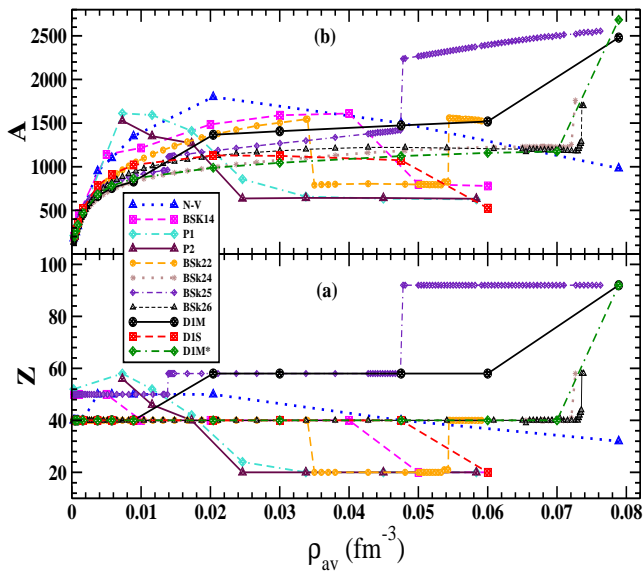}
\caption{\label{fig6b}
	Proton number $Z$ and total number of nucleons $A$ corresponding to 
the $\beta$-equilibrium configuration as a function 
of the inner crust density $\rho_{av}$ for D1M, D1S and D1M* interactions in panels 
(a) and (b), respectively. We also provide the respective numbers predicted by the 
calculation of Negele and Vautherin (N-V) \cite{Negele73}, 
P1, P2 \cite{Baldo07} and BSk22-BSk26 \cite{Pearson18} for comparison.}
\end{figure}
We have seen that when shell effects are added to the semiclassical calculation of the 
energy per particle in WS cells in the inner crust of neutron stars, several local minima 
appear as a function of the proton number considered in the cell. The atomic numbers 
corresponding to these minima depend on the average density of the WS cell and 
change from the standard magic numbers in finite nuclei to the major shell closures 
in a spherical box or in a Woods-Saxon potential {\it without} spin-orbit contribution. 
To get more insight about this evolution of magic numbers predicted 
in WS cells, we consider two specific average densities from the two ends of the 
density range considered, namely $\rho_{\rm av}=0.0004$ fm$^{-3}$ and $\rho_{\rm av}=0.07$ fm$^{-3}$ 
and compute their optimal configurations using the D1M* interaction. From Table 
\ref{tab2} one can see that, for both these average densities the minimum energy 
configuration corresponds to $Z=40$. In panel (a) of Fig. \ref{fig6} we plot the 
proton potentials for the aforementioned average densities. For $\rho_{\rm av}
=0.0004$ fm$^{-3}$ the single-particle proton potential resembles very much to the 
ones corresponding to finite nuclei. The protons seem to be concentrated in a small 
region of the WS cell upto about $\sim 5$ fm. They are affected very little by the 
diluted electron gas, which is smeared throughout the whole WS cell 
(see Fig. \ref{fig6}(c)). For the higher average 
density considered in this example ($\rho_{\rm av}=0.07$ fm$^{-3}$) the radius 
of the WS cell is much smaller, $R_{WS}\sim 16$ fm. This significantly reduces the 
Coulomb effects in the WS cell. In this scenario the proton potential is almost due 
to the nuclear part, which is deep and and almost uniform. It is attenuated from 
$\sim -85$ MeV at the center to $\sim -70$ MeV at the edge, which is quite similar 
to a shallow Woods-Saxon potential with a large radius and diffuseness. The form factors 
of the spin-orbit potential (Eq. (\ref{spinorbitpotential})) are also very different at these two average 
densities considered. This is due to the fact that they 
are determined by the gradients of the neutron and proton densities, which are 
much larger for $\rho_{\rm av}=0.0004$ fm$^{-3}$ compared to those for $\rho_{\rm av}=
0.07$ fm$^{-3}$. These spin-orbit potentials are displayed in the panel (b) of 
Fig. \ref{fig6}. The spin-orbit potential is similar to the cases of finite 
nuclei for the lower average density considered here, however, it is strongly 
damped and shifted outwards for the case of $\rho_{\rm av}=0.07$ fm$^{-3}$, which is in agreement with 
previous findings for the change of the spin-orbit potential near the 
neutron drip-line \cite{Voneiff95,Estal99,Estal01a}. As a consequence of the 
different mean-field $U_p$ and spin-orbit potential $W_p$ entering in the Schr{\"o}dinger 
equation (Eq. (\ref{Schrodinger})), the level schemes are very different for these two 
average densities. For $\rho_{\rm av}=0.0004$ fm$^{-3}$ the magicity appears at 
$Z=20,40,50$ or 82, as in standard nuclei, with quite strong gaps. In the high density 
case, however, as the spin-orbit effect gets much more diluted, the magicity is 
quite similar to the one on a shallow Woods-Saxon potential without spin-orbit with 
major shell closures at $Z=20,40,58,70,90,112$ etc. This is demonstrated in 
Fig. \ref{fig6}(d) and \ref{fig6}(e).

We plot in Fig. \ref{fig6b} the number of protons in panel (a) and total number of 
baryons in panel (b) within the WS cell corresponding to the $\beta$-equilibrium 
configuration as a function of the crustal average density. We compare the results 
obtained with the Gogny forces used in this work with other calculations existing 
in the literature at the WS level. We include in the figure only those calculations 
where the search of the $\beta$-equilibrium configuration consistent with the interaction
used in the calculation has been performed and where the shell corrections have been
explicitly taken into account. 
The number of protons of the selected configurations (see Fig. \ref{fig6b}(a)) lie in the range
 between $Z=20$ and 50 for most of the interactions, with preferences at $Z=20, 40$ or 50. D1M stands out 
among all the interactions considered, having a preference at $Z=58$ for most of the configurations in the
density range considered. The configuration corresponding to 
$\rho_{av}\sim 0.08$ fm$^{-3}$ reaches $Z=92$ for D1M and D1M*, which is quite 
different from the others, though it was also predicted by the calculation using the
BSk25 Skyrme force \cite{Pearson18}. The total number of baryons shown in Fig. \ref{fig6b}(b) depends 
a great deal on the symmetry energy of the respective interactions. This 
is the reason why even with the same proton numbers, different interactions predict 
different neutron content. We want to mention here that there exist in the literature some other studies
of the inner crust within the WS approximation using Skyrme \cite{Grill11,Pastore11}
or Gogny \cite{Than11} interactions. However, these works, mainly devoted to the study
of the pairing properties in the inner crust, use the configurations obtained by
Negele and Vautherin  \cite{Negele73} without searching for the $\beta$-equilibrated
configurations associated with the force.

\begin{figure}[]{}
\includegraphics[height=3.5in,width=3.2in,angle=-90]{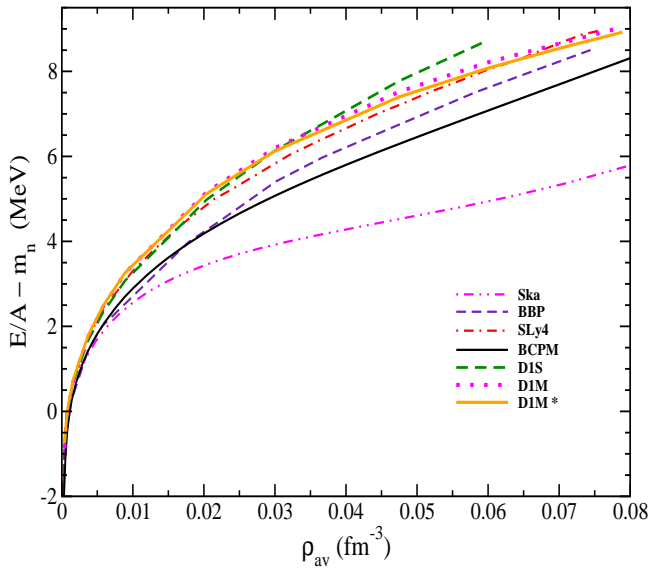}
\caption{\label{fig7}Energy per particle for inner crust of neutron star 
subtracted by free nucleon mass	as a function of density for D1M, D1S and D1M* Gogny forces along 
	with few very successful interactions like BBP, Sly4, BCPM and Ska.}
\end{figure}
\begin{figure}[]{}
\includegraphics[height=3.5in,width=3.2in,angle=-90]{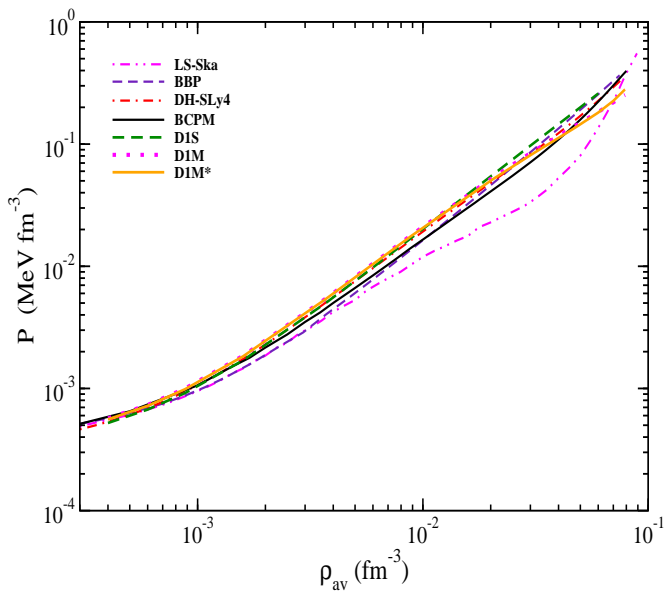}
\caption{\label{fig8}Pressure as a function of density for the inner crust 
	of neutron star using the same set of interactions as Fig. \ref{fig7}.}
\end{figure}
In Figs. \ref{fig7} and \ref{fig8}, we plot the energy per particle subtracted by 
the nucleon mass and the pressure, respectively, as a function of 
the density in the range relevant for the inner crust 
of neutron stars for the Gogny D1S, D1M and D1M* interactions. For comparison we also 
provide the same quantities obtained with the Compressible Liquid 
Drop Model of Baym-Bethe-Pethick (BBP)\cite{Baym71a}, the Skyrme Sly4 \cite{Chabanat98} 
and Ska \cite{Lattimer91a} interactions and with the BCPM 
energy density functional \cite{Baldo08,Baldo10,Baldo13}, which is derived using 
a microscopic interaction from Brueckner calculations with two and three-body forces 
supplemented by a phenomenological term. The energy per particle 
seems to be on the higher side of the figure for all the Gogny interactions compared 
to the test calculations. Specially at higher densities of the 
inner crust, D1S predicts higher energies, whereas D1M and D1M* 
come down in the regime of the Sly4 interaction. 
As far as the pressure is concerned (Fig. \ref{fig8}), 
at higher densities D1M and D1M* have a lowering trend compared to 
others. On the contrary, D1S produces higher pressure. 
The numerical values of the energy per particle subtracted by the nucleon 
mass, the pressure as well as the neutron, proton and electron chemical potentials 
computed with the D1S, D1M and D1M* Gogny forces are also reported in Table \ref{tab2}. 
We want to mention here that, if we plot the pressure as a function of density 
for the Gogny forces computed semiclassically with TF or VWK method, the results are almost 
indistinguishable from the ones we have plotted in Fig. \ref{fig8} obtained with 
VWKSP method, which also includes shell and pairing corrections. 
This points out to the fact that the composition of the inner crust 
does not play a significant role in the equation of state of the neutron star, 
which in turn determines its global properties. 
Out of the three interactions considered here, only D1M* can predict 
neutron stars with masses of 2$M_{\odot}$. We provide the combined EoS of 
core and inner crust calculated with the D1M* interaction in Appendix 
\ref{appendixc}.

\section{Summary and Conclusions}\label{summary}
In summary, we have constructed the EoS of the inner crust of neutron stars using 
the finite-range Gogny forces D1S, D1M and D1M*. For this purpose, we have implemented 
the semiclassical Variational Wigner-Kirkwood method in the spherical 
Wigner-Seitz approximation. Further, we use the Strutinsky integral method 
to add perturbatively the effects of the quantal shell corrections for protons. Pairing correlations are 
added in the BCS approach with the same Gogny force as the mean field.
Details about the theory used in this work are provided in Appendices \ref{appendix1} 
and \ref{appendixb}.

It is found that the quantal effects play a significant role to determine the 
specific composition of the inner crust of the neutron star. We have 
seen that in the inner crust of neutron stars the usual shell 
structures of terrestrial nuclei get washed away at higher 
average densities, where shell closures are similar to the ones of a 
Woods-Saxon potential well without the spin-orbit term. 
In contrast to the composition, we have noticed that the equation of state 
(pressure-vs-density relation) of the inner crust does not get 
much influenced by the shell and pairing effects in the inner crust. Therefore, 
the global properties of the star such as the mass and radius do not get affected 
either. However, for low-mass neutron stars (below the canonical mass of 1.4$M_{\odot}$), 
the stellar radius can change by a considerable amount depending on the 
treatment of the inner crust of neutron stars 
\cite{Baldo14a}, which points out the importance of describing the crust and the core 
with the same interaction. We have compared our results for the energy and pressure
with the ones provided by some popular models of the inner crust available 
in the literature. In our calculations, special attention is paid to the D1M* interaction, which was 
proposed for astrophysical calculations. 
We have obtained a unified EoS for inner crust and core in 
Appendix \ref{appendixc}, which can be used for astrophysical simulations using the D1M* force.

\section{Acknowledgments}
C.M., M.C. and X.V. were partially supported by Grant FIS2017-87534-P from MINECO and FEDER,
and Project MDM-2014-0369 of ICCUB (Unidad de Excelencia Mar\'{\i}a de Maeztu) from MINECO. 
J. N. D. acknowledges support from the Department of Science and Technology
through grant no. EMR/2016/001512.

\section*{APPENDICES}
\appendix
\section{The Variational Wigner-Kirkwood method with shell and paring corrections}\label{appendix1}
The quasilocal energy density functional theory for a finite-range force, established in 
Ref. \cite{Soubbotin03} (also see Refs. \cite{Gridnev17,Behera13,Behera16}),
allows one to write the energy density in a quasilocal form as
\begin{eqnarray}\label{hamiltonian}
\mathcal{H}\equiv\mathcal{H}(\rho_n,\rho_p,\tau_n,\tau_p,\vec{J}_n,
\vec{J}_p),
\end{eqnarray}
where the local particle, kinetic energy and spin densities entering in (\ref{hamiltonian}) 
are obtained in the spirit of the Kohn-Sham scheme from a Slater determinant wave function of single-particle orbitals 
$\phi_i$ as 
\begin{eqnarray}\label{densities}
\rho_q(\vec{r})=\sum_{i=1}^{A_q}\sum_{\sigma}\mid{\phi_i(\vec{r},\sigma,q)}
\mid^2,\nonumber \\
\tau_q(\vec{r})=\sum_{i=1}^{A_q}\sum_{\sigma}\mid{\vec{\nabla}\phi_i(\vec{r},
\sigma,q)}\mid^2,\nonumber \\
{\vec{J}}(\vec{r})=i\sum_{i=1}^{A_q}\sum_{\sigma\sigma^{\prime}}
\phi^*(\vec{r},\sigma,q)\left[(\vec{\sigma})_{\sigma\sigma^{\prime}}
\times\vec{\nabla}\right]\phi_i(\vec{r},\sigma,q).
\end{eqnarray}
The orbitals $\phi_i$ that determine these densities (\ref{densities}) are 
the solutions of the single-particle equations 
\begin{eqnarray}\label{schrodinger}
{h}\phi_i=\left\{-\vec{\nabla}\frac{\hbar^2}{2m_q^*(\vec{r})}
\vec{\nabla}+U_q(\vec{r})-i\vec{W}_q(\vec{r})(\vec{\nabla}\times\vec{\sigma})
\right\}\phi_i=\epsilon_i\phi_i.\nonumber \\
\end{eqnarray}
The effective mass $m_q^*$, the mean-field $U_q$ and spin-orbit
potential $\vec{W}_q$ in (\ref{schrodinger}) are defined as 
\begin{eqnarray}\label{potentials}
\frac{\hbar^2}{2m_q^*}=\frac{\delta \mathcal{H}}{\delta \tau_q},\ \ 
U_q=\frac{\delta \mathcal{H}}{\delta \rho_q},\ \ 
\vec{W}_q=\frac{\delta \mathcal{H}}{\delta \vec{J}_q}.
\end{eqnarray}
They are computed from the energy density (\ref{hamiltonian}) with the
definitions (\ref{densities}) by applying 
the variational principle to the single particle orbitals $\phi_i$. 

In the case of the Gogny interaction (\ref{VGogny}) the quasilocal 
energy density (\ref{hamiltonian}) can be written as
\begin{widetext}
\begin{eqnarray}\label{hamiltonianGogny}
\mathcal{H}=\frac{\hbar^2}{2m}(\tau_n+\tau_p)+\mathcal{H}_{dir}+
\mathcal{H}_{exch}+\mathcal{H}_{zr}
+\mathcal{H}_{coul}+\mathcal{H}_{SO},
\end{eqnarray}
where the different contributions are given by 
\begin{eqnarray}\label{contributions}
&&\mathcal{H}_{dir}=\frac{1}{2}\sum_{i=1}^2\int d\vec{r^{\prime}}\left\{\left(
W_i+\frac{B_i}{2}\right)\rho(\vec{r})\rho(\vec{r^{\prime}})-
\left(H_i+
\frac{M_i}{2}\right)\left[\rho_n(\vec{r})\rho_n(\vec{r^{\prime}})+
\rho_p(\vec{r})\rho_p(\vec{r^{\prime}})\right] \right\}e^{-\frac
{(\vec{r}-\vec{r^{\prime}})^2}{\mu_i^2}},\\
&&\mathcal{H}_{zr}=\frac{t_3}{4} \rho^{\alpha}(\vec{r})\left[
(2+x_3)\rho^2(\vec{r})-(2x_3+1)\right.
\left.(\rho_n^2(\vec{r})+\rho_p^2(\vec{r}))\right],\\
&&\mathcal{H}_{coul}=\frac{1}{2}\int d\vec{r^{\prime}}\frac{\rho_p
(\vec{r})\rho_p(\vec{r^{\prime}})}{\mid\vec{r}-\vec{r^{\prime}}\mid}-
\frac{3}{4}\left(\frac{3}{\pi}\right)^{1\over 3}\rho_p^{4\over 3}(\vec{r}),\\
&&\mathcal{H}_{SO}=-\frac{1}{2}W_0\left[\rho(\vec{r})\vec{\nabla}\cdot
\vec{J}+\rho_n(\vec{r})\vec{\nabla}\cdot\vec{J_n}\right.
\left.+\rho_p(\vec{r})\vec{\nabla}\cdot\vec{J_p}\right].
\label{spinorbitenergy}
\end{eqnarray}
\end{widetext}
Here, $\rho=\rho_n+\rho_p$ and $\vec{J}=\vec{J_n}+\vec{J_p}$ are the 
total particle and spin densities, respectively.

In the quasilocal reduction of the energy density we shall write the exchange contribution in a local form.
To this end, we have to do some approximation to the one-body density matrix, similar to those performed in 
Refs. \cite{Negele72,Negele75,Campi78,Campi79}. In this work we use the Extended Thomas-Fermi density matrix 
derived in Ref. \cite{Soubbotin00}, which has been applied to compute the quantal energy of finite nuclei in the
quasilocal approximation in Refs. \cite{Soubbotin03,Gridnev17}. Using this approximation we can write the 
exchange energy density as a sum of the Thomas-Fermi (Slater) term 
\begin{widetext}
\begin{eqnarray}\label{exchangelocal}
\mathcal{H}_{exch,0}&=&-\frac{1}{2}\sum_i\int d\vec{s} e^{-\frac{s^2}{\mu_i^2}}
\left\{\left(B_i+\frac{W_i}{2}\right)\right.
\left.\left[\left(\rho_n(\vec{r})\frac{3j_1(k_{F_n}s)}{k_{F_n}s}
\right)^2+\left(\rho_p(\vec{r})\frac{3j_1(k_{F_p}s)}{k_{F_p}s}\right)^2\right]
\right.
\nonumber\\
&&\left.-\left(M_i+\frac{H_i}{2}\right)\rho_n(\vec{r})\frac{3j_1(k_{F_n}s)}
{k_{F_n}s}\rho_p(\vec{r})\frac{3j_1(k_{F_p}s)}{k_{F_p}s}\right\},
\end{eqnarray}
\end{widetext}
where $k_{F_q}=(3\pi^2\rho_q(\vec{r}))^{1/3}$ is the local Fermi momentum for
each type of nucleon and 
$j_1$ is the spherical Bessel function, plus a corrective $\hbar^2$-contribution, 
which reads
\begin{widetext}
\begin{eqnarray}\label{exchangelocalh2}
\mathcal{H}_{exch,2}&=&\sum_q\frac{\hbar^2}{2m_q}\left\{(f_q-1)\left(
\tau_q-\frac{3}{5}k_{F_q}^2\rho_q-\frac{1}{4}\Delta\rho_q\right)\right.
\left.+k_{F_q}f_{qk}\left[\frac{1}{27}\frac{(\vec{\nabla}\rho_q)^2}
{\rho_q}-\frac{1}{36}\Delta\rho_q\right]\right\}.
\end{eqnarray}
\end{widetext}
In this equation 
\begin{eqnarray}\label{definitionf}
	f_q\equiv f_q(\vec{r},k)_{k=k_{F_q}}\ \ \text{and}\ \ 
f_{qk}\equiv \left(\frac{\partial f(\vec{r},k)}{\partial k}\right)_{k=k_{F_q}}
\end{eqnarray}
are the 
inverse of the position and momentum-dependent effective mass and its 
derivative with respect to the momentum, both computed at the corresponding 
local Fermi momentum for each kind of nucleons.  
The quantity $f_q(\vec{r},k)$ that enters in Eq. \ref{exchangelocalh2} is defined as
\begin{eqnarray}\label{fq}
f_q(\vec{r},k)=1+\frac{m}{\hbar^2 k}\frac{\partial U_{exch,q}(\vec{r},k)}
	{\partial k},
\end{eqnarray}
where $U_{exch,q}$ is the Wigner transform of the single-particle exchange 
potential in the TF approximation, which can be written as  
\begin{widetext}
\begin{eqnarray}\label{vexch}
U_{exch,q}(\vec{r},k)&=&-\sum_i\int d\vec{s}e^{-i\vec{k}\cdot\vec{s}}
e^{-\frac{s^2}{\mu_i^2}}\left\{\left(B_i+\frac{W_i}{2}\right)\rho_q
(\vec{r})\right.
\left.\frac{3j_1(k_{F_q}s)}{k_{F_q}s}\right.
\nonumber \\
&-&\left.\left(M_i+\frac{H_i}{2}
\right)\left[\rho_q(\vec{r})\frac{3j_1(k_{F_q}s)}{k_{F_q}s}\right.
\left.+\rho_{q^{\prime}}(\vec{r})\frac{3j_1(k_{F_{q^\prime}}s)}
{k_{F_{q^\prime}}s}\right]\right\},
\end{eqnarray}
\end{widetext}
where, $q=n,p$ and $q^{\prime}=p,n$. Notice that the exchange potential at TF level
is a function not only of the momentum $k$ of the nucleon of type $q$, but also   
of the position via the dependence of the TF exchange potential on the local Fermi
momentum of both, neutrons and protons, $k_{F_n}(\vec{r})$ and $k_{F_p}(\vec{r})$, 
respectively.
 
Combining the total kinetic energy density with the $\hbar^2$ part of the exchange energy density
(\ref{exchangelocalh2}), one can sort out explicitly the effective mass contribution,
which at pure TF level is hidden in the exchange term. In this way we can write
\begin{widetext}
\begin{eqnarray}\label{h2energy2}
\widetilde{\mathcal E}=\sum_q\left[\frac{\hbar^2}{2m}f_q\tau_q+\frac{\hbar^2}{2m}\left\{
\left(1-f_q\right)\tau_{q,0}-\frac{1}{4}f_q\Delta\rho_q
+k_{F_q}f_{qk}\left(
\frac{1}{27}\frac{\left(\vec{\nabla}\rho_q\right)^2}{\rho_q}-\frac{1}{36}\Delta\rho_q
\right)\right\}\right],
\end{eqnarray}
\end{widetext}
where $\tau_q=\tau_{q,0}+\tau_{q,2}$ contains the pure TF and $\hbar^2$ 
contributions. 

In the $\hbar^2$ contribution to the exchange energy 
(\ref{exchangelocalh2}), $\tau_q$ is the semiclassical 
kinetic energy density for each type of particles, which is  obtained from the 
semiclassical ETF density matrix \cite{Soubbotin00,Gridnev17} as
\begin{widetext}
\begin{eqnarray}\label{tauetf}
&&\tau_q(\vec{r})=\left.\left(\frac{1}{4}\Delta_R-\Delta_s\right)
\widetilde{\rho_q}(\vec{R},s)\right|_{s=0}=\frac{3}{5}k_{F_q}^2 s_q
+\frac{1}{36}\frac{(\vec{\nabla}\rho_q)^2}{\rho_q}\left[1+\frac{2}{3}
k_{F_q}\frac{f_{qk}}{f_q}+\frac{2}{3}k_{F_q}^2\frac{f_{qkk}}{f_q}-
\frac{1}{3}k_{F_q}^2\frac{f_{qk}^2}{f_q^2}\right]\nonumber\\
&&+\frac{1}{4}\Delta\rho\left[1+\frac{2}{9}k_{F_q}\frac{f_{qk}}{f_q}\right]
+\frac{1}{6}\rho_q\frac{\Delta f_{q}}{f_q}+\frac{1}{6}\frac{\vec{\nabla}
\rho_q\cdot\vec{\nabla}f_q}{f_q}
\left[1-\frac{1}{3}k_{F_q}\frac{f_{qk}}{f_q}\right]+\frac{1}{9}
\frac{\vec{\nabla}\rho_q\cdot\vec{\nabla}f_{qk}}{f_q}-\frac{1}{12}\rho_q
\frac{(\vec{\nabla}f_q)^2}{f_q^2},
\end{eqnarray}
\end{widetext}
which is a functional of the two kind of local densities $\rho_q$ and  $\rho_{q'}$,
the latest owing to the presence of effective mass terms in (\ref{tauetf}). Therefore, 
including $\hbar^2$ corrections the energy density (\ref{hamiltonianGogny}) becomes a functional of the particle densities only. 

As explained in Ref. \cite{Gridnev17}, 
the spin-orbit interaction provides an additional term to the semiclassical 
ETF density matrix, which allows one to calculate the semiclassical spin density 
as 
\begin{widetext}
\begin{eqnarray}\label{spindensity}
	\vec{J}_q(\vec{R})=-i \text{Tr}
\left\{\left[\vec{\sigma}\times\left(\frac{\vec{\nabla}_R}{2}+\vec{\nabla}_s
\right)\right]
\left(-\frac{im}{2\hbar^2}\right)\frac{\rho_q}{f_q}\vec{\sigma}
\cdot(\vec{W}_q\times\vec{s})\frac{3j_1(k_{F_q}s)}{k_{F_q}s}\right\}_{s=0}
= -\frac{2m}{\hbar^2}\frac{\rho_q \vec{W}_q}{f_q},
\end{eqnarray}
\end{widetext}
where, the spin-orbit potential $\vec{W}_q$ (see third equation in (\ref{potentials}))
is given by 
\begin{eqnarray}\label{spinorbitpotential}
\vec{W}_q(\vec{r})=\frac{1}{2}W_0\left[\vec{\nabla}\rho+\vec{\nabla}\rho_q
\right].
\end{eqnarray}
The spin-orbit term in the density matrix also provides another 
contribution to the kinetic energy density (\ref{tauetf}) given by 
\begin{eqnarray}\label{tauso}
\tau_{q,SO}=\frac{1}{2}\left(\frac{2m}{\hbar^2}\right)^2\frac{\rho}{f_q^2}
W_q^2.
\end{eqnarray}
Using the spin-density (\ref{spindensity}) in the spin-orbit energy 
density (\ref{spinorbitenergy}) and performing a suitable partial 
integration, one can write the semiclassical spin-orbit energy density,
which is actually a $\hbar^2$ order quantity, as
\begin{eqnarray}\label{spinorbitenergyetf}
\mathcal{H}_{SO}=\vec{J}_n\cdot\vec{W}_n+\vec{J}_p\cdot\vec{W}_p
=-\frac{2m}{\hbar^2}\left[\frac{\rho_n W_n^2}{f_n}+\frac{\rho_p W_p^2}{f_p}
\right].\ \ \ \     
\end{eqnarray}

\section{The single-particle potential}\label{appendixb}
In order to describe the quantal energy of a nucleus using a finite-range
interaction within a Mic-Mac frame, we shall add perturbatively
to the macroscopic part given by the semiclassical energy the shell and pairing corrections. 
For each type of particles these quantal effects are obtained starting from the mean field 
obtained semiclassically by performing the variation of the energy density (\ref{hamiltonianGogny}) with respect to
the neutron or proton densities (see second equation (\ref{potentials})). The $\hbar^0$ (TF)
part of the single-particle potential consists of the direct and  zero-range contributions, 
which read
\begin{widetext}
\begin{eqnarray}\label{meanfield}
&&U_{dir,q}=\sum_{i=1}^2\int d\vec{r^{\prime}}\left\{\left(
W_i+\frac{B_i}{2}\right)\rho(\vec{r^{\prime}})-
\left(H_i+\frac{M_i}{2}\right)\rho_n(\vec{r^{\prime}})\right\}e^{-\frac
{(\vec{r}-\vec{r^{\prime}})^2}{\mu_i^2}},\\
&&U_{zr,q}=\frac{t_3}{2} \rho^{\alpha}(\vec{r})\left[
(2+x_3)\rho(\vec{r})-(2x_3+1)\rho_q(\vec{r})\right]
+ \frac{t_3\alpha}{2} \rho^{\alpha-1}(\vec{r})\left[
(2+x_3)\rho^2(\vec{r})-(2x_3+1)(\rho_n^2(\vec{r})+\rho_p^2(\vec{r}))\right],
\end{eqnarray}
\end{widetext}
plus the exchange term given by Eq.(\ref{exchangelocal}), i.e.,
\begin{eqnarray}
U_{q,0} = U_{dir,q}+U_{exch,q}+U_{zr,q}
\end{eqnarray}
and, in the case of protons, also including the contribution of the Coulomb potential
\begin{eqnarray}
U_{coul}=\int d\vec{r^{\prime}}\frac{\rho_p(\vec{r^{\prime}})}
{\mid\vec{r}-\vec{r^{\prime}}\mid}-
\left(\frac{3}{\pi}\right)^{1\over 3}\rho_p^{1\over 3}(\vec{r}).
\label{coulomb}
\end{eqnarray}

Next we compute the $\hbar^2$ contribution to the single-particle
potential. Notice that the spin-orbit energy (\ref{spinorbitenergy}) is 
also a $\hbar^2$-order quantity that depends on the particle and spin 
densities for each type of particles. As mentioned before, the neutron 
and proton spin-orbit potentials, defined by Eq. (\ref{spinorbitpotential}), 
come from the variation of Eq. (\ref{spinorbitenergy}) with respect to the 
spin densities. The spin-orbit energy in Eq. (\ref{spinorbitenergy}) also 
contributes to the mean field of each type of particles through its 
variation with respect to the corresponding particle density, which is 
given as
\begin{eqnarray}
&&U_{SO}=-\frac{1}{2}W_0\left[\vec{\nabla}\cdot
\vec{J}(\vec{r})+\vec{\nabla}\cdot\vec{J}_q(\vec{r})\right].
\label{spinorbitmf}
\end{eqnarray}
Therefore, in order to include the spin-orbit contributions in our formalism, 
which are essential for describing properly the shell effects through the 
Strutinsky integral method, the semiclassical expansion of the energy should 
be pushed, at least, up to $\hbar^2$-order.

To obtain the nuclear part of the $\hbar^2$ contribution to the single-particle potential, 
one needs to perform the variation of the combination of the kinetic and second-order 
exchange energy densities $\widetilde{\mathcal E}$ (\ref{h2energy2}) with respect to 
the neutron or proton densities $\rho_q$. To perform this variation one needs to 
treat the $\tau_q$ as an independent variable in $\widetilde{\mathcal E}$. So, this 
$\hbar^2$ contribution to the single-particle potential is given by
\begin{widetext}
\begin{eqnarray}\label{h2pot1}
&&U_{q,2}=\frac{\delta \widetilde{\mathcal E}}{\delta \rho_q}=\frac{\hbar^2}{2m}
\left(\frac{\delta f_q}{\delta \rho_q}\tau_q+\frac{\delta f_{q^\prime}}{\delta \rho_q}\tau_{q^\prime}\right)
+\frac{\hbar^2}{2m}\frac{\delta }{\delta \rho_q}\left\{(1-f_q)\tau_{q,0}+(1-f_{q^\prime})\tau_{q^\prime,0}\right\}\nonumber\\
&+&\frac{\hbar^2}{2m}\frac{\delta }{\delta \rho_q}\left\{-\frac{1}{4}f_q\Delta\rho_q-\frac{1}{4}f_{q^\prime}\Delta\rho_{q^\prime}
+k_{F_q}f_{qk}\left(\frac{1}{27}\frac{\left(\vec{\nabla}\rho_q\right)^2}{\rho_q}
-\frac{1}{36}\Delta\rho_q\right)+k_{F_{q^\prime}}f_{q^\prime k}\left(\frac{1}{27}
\frac{\left(\vec{\nabla}\rho_{q^\prime}\right)^2}{\rho_{q^\prime}}
-\frac{1}{36}\Delta\rho_{q^\prime}\right)\right\}.\ \ \ \ \
\end{eqnarray}
Now, with the definitions
$f_q\equiv f_q(k=k_{F_q},k_{F_q},k_{F_{q^\prime}})\ \text{and}\ 
f_{q^\prime}\equiv f_{q^\prime}(k=k_{F_{q^\prime}},k_{F_{q^\prime}},k_{F_q})$,
one can easily obtain
\begin{eqnarray}\label{fq2}
 \frac{\partial f_q}{\partial \rho_q}=\frac{\partial f_q}{\partial k_{F_q}} 
\frac{\partial k_{F_q}}{\partial \rho_q}=\left(f_{qk}+f_{qk_{F_q}}\right)
\frac{1}{3}\frac{k_{F_q}}{\rho_q}\ \text{and}\ 
	\frac{\partial f_{q^\prime}}{\partial \rho_q}=f_{q^\prime k_{F_q}}
	\frac{1}{3}\frac{k_{F_q}}{\rho_q}.
\end{eqnarray}
	Using this the variation of the first term in Eq. (\ref{h2pot1}) is given by, 
\begin{eqnarray}\label{variation2}
\frac{\hbar^2}{2m}\left(\frac{\delta f_q}{\delta \rho_q}\tau_q+\frac{\delta f_{q^\prime}}{\delta \rho_q}\tau_{q^\prime}\right)
=\frac{\hbar^2}{2m}\frac{1}{3}\frac{k_{F_q}}{\rho_q}\left\{\left(f_{qk}+f_{qk_{F_q}}\right)\tau_q
+f_{q^\prime k_{F_q}}\tau_{q^\prime}\right\}.
\end{eqnarray}
Now taking into account that, $\tau_{q,0}=\frac{3}{5}(3\pi^2)^{2/3}\rho_q^{5/3}$, one can write 
after some algebraic simplifications,
\begin{eqnarray}\label{variation1}
\frac{\hbar^2}{2m}\frac{\delta }{\delta \rho_q}
\left\{(1-f_q)\tau_{q,0}+(1-f_{q^\prime})\tau_{q^\prime,0}\right\}
=\frac{\hbar^2}{2m}\frac{1}{3\rho_q}\left[5\left\{1-f_q-\frac{1}{5}k_{F_q}\left(
f_{qk}+f_{qk_{F_q}}\right)\right\}\tau_{q,0}-f_{q^\prime k_{F_q}}k_{F_q}{\tau_{q^\prime,0}}\right].
\end{eqnarray}
Combining Eq. (\ref{variation1}) and (\ref{variation2}) one gets,
\begin{eqnarray}\label{variation3}
\frac{\hbar^2}{2m}\left[\frac{\delta f_q}{\delta \rho_q}\tau_q+\frac{\delta f_{q^\prime}}{\delta \rho_q}\tau_{q^\prime}
+(1-f_q)\tau_{q,0}+(1-f_{q^\prime})\tau_{q^\prime,0}\right]=\frac{\hbar^2}{2m}\frac{1}{3\rho_q}\left[5\left\{1-f_q\right\}
	\tau_{q,0}+k_{F_q}\left(f_{qk}+f_{qk_{F_q}}\right)\tau_{q,2}+k_{F_q}f_{q^\prime k_{F_q}}\tau_{q^\prime,2}\right],\nonumber\\
\end{eqnarray}
\end{widetext}
where in this equation we use the local kinetic energy density 
$\tau_{q,2}=\frac{1}{36}\frac{(\vec{\nabla}\rho_q)^2}{\rho_q}+\frac{1}{3}
\Delta \rho_q$. As explained in Refs. \cite{Soubbotin00} and 
\cite{Gridnev17}, one obtains almost the same $\hbar^2$-order energy if in the 
full kinetic energy density (\ref{variation3}) the $\hbar^2$ contribution 
is replaced by its local counterpart. 

The contributions of the remaining pieces of Eq.(\ref{h2pot1}), which correspond to the $\hbar^2$-part of the
single particle potential are given after some algebraic steps by, 
\begin{widetext}
\begin{eqnarray}\label{h2pot2}
&&\frac{\hbar^2}{2m}\frac{\delta}{\delta \rho_q}
\left\{-\frac{1}{4}f_q\Delta\rho_q-\frac{1}{4}f_{q^\prime}\Delta\rho_{q^\prime}
+k_{F_q}f_{qk}\left(\frac{1}{27}\frac{\left(\vec{\nabla}\rho_q\right)^2}{\rho_q}
-\frac{1}{36}\Delta\rho_q\right)+k_{F_{q^\prime}}f_{q^\prime k}\left(\frac{1}{27}
\frac{\left(\vec{\nabla}\rho_{q^\prime}\right)^2}{\rho_{q^\prime}}
-\frac{1}{36}\Delta\rho_{q^\prime}\right)\right\}\nonumber\\
&=&\Bigg\{\left[2k_{F_q}\left(f_{qk}+f_{qk_{F_q}}\right)-k_{F_q}^2
\left(f_{qkk}+2f_{qkk_{F_q}}+f_{qk_{F_q}k_{F_q}}\right)\right]
+\frac{1}{324}\left[10k_{F_q}f_{qk}-4k_{F_q}^2\left(f_{qkk}+f_{qkk_{F_q}}\right)\right.\nonumber\\
&-&\left. k_{F_q}^3\left(f_{qkkk}+2f_{qkkk_{F_q}}+f_{qkk_{F_q}k_{F_q}}\right)\right]\Bigg\}
\frac{(\vec{\nabla}\rho_q)^2}{\rho_q^2}\nonumber\\
&-&\left\{\frac{1}{6}k_{F_q}\left(f_{qk}+f_{qk_{F_q}}\right)+\frac{1}{108}\left[10k_{F_q}f_{qk}
+2k_{F_q}^2\left(f_{qkk}+f_{qkk_{F_q}}\right)\right]\right\}\frac{\Delta \rho_q}{\rho_q}\nonumber\\
&-&\frac{1}{162}\left[14k_{F_q}k_{F_{q^\prime}}f_{qkk_{F_{q^\prime}}}+k_{F_q}^2 k_{F_{q^\prime}}
f_{qkkk_{F_{q^\prime}}}\right]\frac{\vec{\nabla}\rho_q}{\rho_q}\frac{\vec{\nabla}\rho_{q^\prime}}
{\rho_{q^\prime}}\nonumber\\
&+&\frac{1}{324}\left[4k_{F_q}k_{F_{q^\prime}}f_{q^\prime kk_{F_q}}\frac{\rho_{q^\prime}}
{\rho_q}+18k_{F_{q^\prime}}f_{qk_{F_{q^\prime}}}-9k_{F_{q^\prime}}^2f_{qk_{F_{q^\prime}}}k_{F_{q^\prime}}
+ 2k_{F_q}k_{F_{q^\prime}}f_{qkk_{F_{q^\prime}}}-k_{F_{q^\prime}}^2k_{F_q}f_{qkk_{F_{q^\prime}}k_{F_{q^\prime}}}
\right]\frac{(\vec{\nabla}\rho_{q^\prime})^2}{\rho_{q^\prime}^2}\nonumber\\
&-&\frac{1}{108}\left[\left(9
k_{F_q}f_{q^\prime k_{F_q}}+k_{F_q}k_{F_{q^\prime}}f_{q^\prime kk_{F_q}}\right)\frac{\rho_{q^\prime}}{\rho_q}
+9k_{F_{q^\prime}}f_{qk_{F_{q^\prime}}}+k_{F_q}k_{F_{q^\prime}}f_{qkk_{F_{q^\prime}}}\right]
\frac{\Delta \rho_{q^\prime}}{\rho_{q^\prime}}.
\end{eqnarray}
\end{widetext}

Here, we have used under the integral sign the rules of functional derivative as, 
\begin{eqnarray}\label{functionalder}
\frac{\delta F[\rho]}{\delta \rho}=\frac{\partial f}{\partial \rho}+\sum_{i=1}^N (-1)^i\nabla^{(i)}
\frac{\partial f}{\partial (\nabla^{(i)}\rho)},
\end{eqnarray}
where the functional $F[\rho(r)]$ is defined as, 
\begin{eqnarray}\label{functional}
F[\rho(r)]=\int f(\vec{r},\rho(\vec{r}),\vec{\nabla}\rho(\vec{r}),\vec{\nabla}^{(2)}\rho(\vec{r})\cdots) d\vec{r}.
\end{eqnarray}

Now, to obtain the 
binding energy of a set of nuclei including shell and pairing effects, 
as close as possible to the full HF or HFB values overall, a scaling 
parameter $\beta_{\scriptscriptstyle KE}$ has been introduced in the $\hbar^2$-part of the kinetic 
energy density $\tau_q$. Incorporating this parameter, the kinetic energy density now becomes,
\begin{eqnarray}\label{kebetakt}
{\mathcal H}_{KE}=\sum_q \frac{\hbar^2}{2m} f_q(\tau_{q,0}+\beta_{\scriptscriptstyle KE}\tau_{q,2}),
\end{eqnarray}
and the potential part corresponding to the $\hbar^2$ term given in Eq. (\ref{variation3})
containing $\tau_{q,2}$ and $\tau_{q^\prime,2}$ is modified as $\beta_{\scriptscriptstyle KE}\tau_{q,2}$ 
and $\beta_{\scriptscriptstyle KE}\tau_{q^\prime,2}$ respectively. Following this procedure, 
we find $\beta_{\scriptscriptstyle KE}=1.45$, 1.4 and 1.75 for D1S, D1M and D1M* interactions,
respectively.

\section{EoS for D1M*}\label{appendixc}
\squeezetable
\begin{table*}[h]
        \caption{Equation of state for inner crust and core of NS obtained with D1M* interaction. Here 
	we have used $\mathcal{E}=\frac{E}{A}\cdot \rho$. $X_p$, $X_e$ and $X_{\mu}$ are proton, electron 
	and muon fraction in the medium respectively.}
  \label{tab3}
\begin{tabular}{ccccccccc}
\toprule
	&$\rho$ & $E/A-m_n$ & $P$ &  $\mathcal{E}$ & $P$& $X_p$ & $X_e$ & $X_{\mu}$ \\ 
	&(fm$^{-3}$) & (MeV) & (MeV fm$^{-3}$) & (g cm$^{-3}$) & (erg cm$^{-3}$) &  &  &  \\
\hline
	Crust&	0.0004  &  -0.74943  &    0.00056 &  6.6903$\times 10^{11}$      &  8.9472$\times 10^{29}$ &  0.20234 &  0.20234 &   0.00000  \\                                    
	     &  0.0006  &  -0.27892  &    0.00072 &  1.0040$\times 10^{12}$      &  1.1545$\times 10^{30}$ &  0.15285 &  0.15285 &   0.00000  \\                                    
	     &  0.000879&   0.12384  &    0.00098 &  1.4716$\times 10^{12}$      &  1.5760$\times 10^{30}$ &  0.11938 &  0.11938 &   0.00000  \\                                    
	     &  0.00159 &   0.74518  &    0.00183 &  2.6636$\times 10^{12}$      &  2.9277$\times 10^{30}$ &  0.08500 &  0.08500 &   0.00000  \\                                    
	     &  0.00373 &   1.79723  &    0.00543 &  6.2557$\times 10^{12}$      &  8.6995$\times 10^{30}$ &  0.05853 &  0.05853 &   0.00000  \\                                    
	     &  0.00577 &   2.47121  &    0.00983 &  9.6839$\times 10^{12}$      &  1.5749$\times 10^{31}$ &  0.05115 &  0.05115 &   0.00000  \\                                    
	     &  0.00891 &   3.26236  &    0.01770 &  1.4966$\times 10^{13}$      &  2.8359$\times 10^{31}$ &  0.04624 &  0.04624 &   0.00000  \\                                    
	     &  0.0204  &   5.12315  &    0.05114 &  3.4334$\times 10^{13}$      &  8.1938$\times 10^{31}$ &  0.04048 &  0.04048 &   0.00000  \\                                    
	     &  0.03    &   6.12660  &    0.08046 &  5.0545$\times 10^{13}$      &  1.2891$\times 10^{32}$ &  0.03827 &  0.03827 &   0.00000  \\                                    
	     &  0.0475  &   7.40033  &    0.13496 &  8.0138$\times 10^{13}$      &  2.1623$\times 10^{32}$ &  0.03565 &  0.03565 &   0.00000  \\                                    
	     &  0.06    &   8.07662  &    0.17845 &  1.0130$\times 10^{14}$      &  2.8591$\times 10^{32}$ &  0.03453 &  0.03453 &   0.00000  \\                                    
	     &  0.07    &   8.53993  &    0.22176 &  1.1824$\times 10^{14}$      &  3.5529$\times 10^{32}$ &  0.03406 &  0.03406 &   0.00000  \\
	     &  0.0789  &   8.92225  &    0.26701 &  1.3333$\times 10^{14}$      &  4.2779$\times 10^{32}$ &  0.03429 &  0.03429 &   0.00000  \\
\hline
	Core&0.0838&        9.11446  &        0.27042  &    1.4164$\times 10^{14}$  &    4.3326$\times 10^{32}$  &        0.03502  &        0.03502  &        0.00000\\
        &0.09  &        9.35675  &        0.32190  &    1.5215$\times 10^{14}$  &    5.1573$\times 10^{32}$  &        0.03612  &        0.03612  &        0.00000  \\
	&0.10  &        9.76819  &        0.42733  &    1.6913$\times 10^{14}$  &    6.8466$\times 10^{32}$  &        0.03765  &        0.03765  &        0.00000  \\
	&0.11  &       10.21459  &        0.56487  &    1.8613$\times 10^{14}$  &    9.0502$\times 10^{32}$  &        0.03893  &        0.03893  &        0.00000  \\
	&0.12  &       10.70424  &        0.73946  &    2.0316$\times 10^{14}$  &    1.1848$\times 10^{33}$  &        0.04000  &        0.04000  &        0.00000  \\
	&0.13  &       11.23588  &        0.95653  &    2.2021$\times 10^{14}$  &    1.5325$\times 10^{33}$  &        0.04090  &        0.04082  &        0.00008  \\
	&0.14  &       11.77626  &        1.22382  &    2.3729$\times 10^{14}$  &    1.9608$\times 10^{33}$  &        0.04166  &        0.04098  &        0.00068  \\
	&0.15  &       12.36780  &        1.54275  &    2.5440$\times 10^{14}$  &    2.4718$\times 10^{33}$  &        0.04232  &        0.04090  &        0.00141  \\
	&0.16  &       13.02104  &        1.91706  &    2.7154$\times 10^{14}$  &    3.0715$\times 10^{33}$  &        0.04288  &        0.04072  &        0.00216  \\
	&0.17  &       13.73982  &        2.35077  &    2.8873$\times 10^{14}$  &    3.7663$\times 10^{33}$  &        0.04337  &        0.04048  &        0.00289  \\
	&0.18  &       14.52609  &        2.84784  &    3.0597$\times 10^{14}$  &    4.5627$\times 10^{33}$  &        0.04379  &        0.04021  &        0.00358  \\
	&0.21  &       17.29766  &        4.75742  &    3.5800$\times 10^{14}$  &    7.6222$\times 10^{33}$  &        0.04483  &        0.03938  &        0.00545\\
	&0.24  &       20.69063  &        7.36949  &    4.1059$\times 10^{14}$  &    1.1807$\times 10^{34}$  &        0.04567  &        0.03864  &        0.00704\\
	&0.27  &       24.69654  &       10.77234  &    4.6385$\times 10^{14}$  &    1.7259$\times 10^{34}$  &        0.04647  &        0.03804  &        0.00843\\
	&0.30  &       29.29993  &       15.04513  &    5.1785$\times 10^{14}$  &    2.4105$\times 10^{34}$  &        0.04732  &        0.03763  &        0.00969\\
	&0.33  &       34.48178  &       20.25854  &    5.7268$\times 10^{14}$  &    3.2458$\times 10^{34}$  &        0.04830  &        0.03741  &        0.01089\\
	&0.36  &       40.22117  &       26.47535  &    6.2842$\times 10^{14}$  &    4.2418$\times 10^{34}$  &        0.04945  &        0.03739  &        0.01207\\
	&0.39  &       46.49613  &       33.75094  &    6.8515$\times 10^{14}$  &    5.4075$\times 10^{34}$  &        0.05083  &        0.03756  &        0.01327\\
	&0.42  &       53.28414  &       42.13356  &    7.4294$\times 10^{14}$  &    6.7505$\times 10^{34}$  &        0.05246  &        0.03794  &        0.01451\\
	&0.45  &       60.56242  &       51.66452  &    8.0185$\times 10^{14}$  &    8.2776$\times 10^{34}$  &        0.05437  &        0.03853  &        0.01584\\
	&0.48  &       68.30797  &       62.37829  &    8.6193$\times 10^{14}$  &    9.9941$\times 10^{34}$  &        0.05661  &        0.03934  &        0.01727\\
	&0.51  &       76.49773  &       74.30242  &    9.2325$\times 10^{14}$  &    1.1905$\times 10^{35}$  &        0.05920  &        0.04038  &        0.01882\\
	&0.54  &       85.10850  &       87.45746  &    9.8585$\times 10^{14}$  &    1.4012$\times 10^{35}$  &        0.06217  &        0.04165  &        0.02053\\
	&0.57  &       94.11700  &      101.85674  &    1.0498$\times 10^{15}$  &    1.6319$\times 10^{35}$  &        0.06556  &        0.04316  &        0.02240\\
	&0.60  &      103.49989  &      117.50632  &    1.1151$\times 10^{15}$  &    1.8827$\times 10^{35}$  &        0.06940  &        0.04493  &        0.02447\\
	&0.63  &      113.23382  &      134.40496  &    1.1817$\times 10^{15}$  &    2.1534$\times 10^{35}$  &        0.07371  &        0.04697  &        0.02674\\
	&0.66  &      123.29558  &      152.54446  &    1.2499$\times 10^{15}$  &    2.4440$\times 10^{35}$  &        0.07851  &        0.04928  &        0.02923\\
	&0.69  &      133.66233  &      171.91034  &    1.3194$\times 10^{15}$  &    2.7543$\times 10^{35}$  &        0.08381  &        0.05186  &        0.03195\\
	&0.72  &      144.31192  &      192.48320  &    1.3904$\times 10^{15}$  &    3.0839$\times 10^{35}$  &        0.08961  &        0.05470  &        0.03491\\
	&0.75  &      155.22325  &      214.24048  &    1.4630$\times 10^{15}$  &    3.4325$\times 10^{35}$  &        0.09589  &        0.05780  &        0.03809\\
	&0.78  &      166.37673  &      237.15886  &    1.5370$\times 10^{15}$  &    3.7997$\times 10^{35}$  &        0.10262  &        0.06113  &        0.04148\\
	&0.81  &      177.75459  &      261.21683  &    1.6125$\times 10^{15}$  &    4.1852$\times 10^{35}$  &        0.10974  &        0.06467  &        0.04507\\
	&0.84  &      189.34119  &      286.39706  &    1.6896$\times 10^{15}$  &    4.5886$\times 10^{35}$  &        0.11718  &        0.06837  &        0.04881\\
	&0.87  &      201.12315  &      312.68832  &    1.7682$\times 10^{15}$  &    5.0098$\times 10^{35}$  &        0.12488  &        0.07220  &        0.05268\\
	&0.90  &      213.08924  &      340.08655  &    1.8484$\times 10^{15}$  &    5.4488$\times 10^{35}$  &        0.13275  &        0.07612  &        0.05663\\
	&0.93  &      225.23028  &      368.59505  &    1.9301$\times 10^{15}$  &    5.9055$\times 10^{35}$  &        0.14070  &        0.08007  &        0.06063\\
	&0.96  &      237.53882  &      398.22395  &    2.0135$\times 10^{15}$  &    6.3803$\times 10^{35}$  &        0.14868  &        0.08403  &        0.06464\\
	&0.99  &      250.00888  &      428.98910  &    2.0984$\times 10^{15}$  &    6.8732$\times 10^{35}$  &        0.15660  &        0.08796  &        0.06863\\
	&1.02  &      262.63561  &      460.91087  &    2.1850$\times 10^{15}$  &    7.3846$\times 10^{35}$  &        0.16442  &        0.09184  &        0.07258\\
	&1.05  &      275.41506  &      494.01278  &    2.2731$\times 10^{15}$  &    7.9150$\times 10^{35}$  &        0.17208  &        0.09563  &        0.07645\\
	&1.08  &      288.34398  &      528.32039  &    2.3630$\times 10^{15}$  &    8.4646$\times 10^{35}$  &        0.17956  &        0.09933  &        0.08023\\
	&1.11  &      301.41961  &      563.86039  &    2.4545$\times 10^{15}$  &    9.0340$\times 10^{35}$  &        0.18683  &        0.10292  &        0.08391\\
&1.14  &      314.63960  &      600.65982  &    2.5477$\times 10^{15}$  &    9.6236$\times 10^{35}$  &        0.19387  &        0.10639  &        0.08749\\
&1.17  &      328.00187  &      638.74563  &    2.6426$\times 10^{15}$  &    1.0234$\times 10^{36}$  &        0.20067  &        0.10973  &        0.09094\\
&1.20  &      341.50460  &      678.14433  &    2.7392$\times 10^{15}$  &    1.0865$\times 10^{36}$  &        0.20723  &        0.11295  &        0.09428\\
&1.23  &      355.14611  &      718.88173  &    2.8376$\times 10^{15}$  &    1.1518$\times 10^{36}$  &        0.21354  &        0.11605  &        0.09749\\
&1.26  &      368.92487  &      760.98288  &    2.9378$\times 10^{15}$  &    1.2192$\times 10^{36}$  &        0.21961  &        0.11902  &        0.10059\\
&1.29  &      382.83948  &      804.47197  &    3.0397$\times 10^{15}$  &    1.2889$\times 10^{36}$  &        0.22544  &        0.12187  &        0.10356\\
&1.32  &      396.88859  &      849.37238  &    3.1435$\times 10^{15}$  &    1.3608$\times 10^{36}$  &        0.23103  &        0.12460  &        0.10643\\
&1.35  &      411.07097  &      895.70663  &    3.2491$\times 10^{15}$  &    1.4351$\times 10^{36}$  &        0.23640  &        0.12722  &        0.10918\\
&1.38  &      425.38542  &      943.49646  &    3.3565$\times 10^{15}$  &    1.5116$\times 10^{36}$  &        0.24155  &        0.12973  &        0.11182\\
&1.41  &      439.83081  &      992.76285  &    3.4658$\times 10^{15}$  &    1.5906$\times 10^{36}$  &        0.24648  &        0.13213  &        0.11435\\
\toprule
 \end {tabular}
\end{table*}


\end{document}